\documentstyle[preprint,aps]{revtex}

\begin{document}
\author{ M.E. PALISTRANT}

\title{SUPERCONDUCTIVITY IN TWO-BAND SYSTEM WITH LOW CARRIER DENSITY}
\maketitle

\begin{center}
{\it Institute of Applied Physics, Academy of Sciences of Moldova,\\
5 Akademiei Street, 2028 Chishinau, Moldova}
\end{center}

\begin{center}
\sf{Received 25 octouber 2003}
\end{center}

\begin{quotation}

\sf{In the article a review over the theory of superconductors with energy bands
that overlap on Fermi surface at arbitrary densities of charge carriers
(including reduced and very low) is done. All pairings of electrons that
result in formation Cooper pairs of electrons from different energy bands as
well as in every energy band are considered.

The system of equations for four order parameters $\Delta _{11}$, $\Delta
_{22}$, $\Delta _{12}$, $\Delta _{21}$ and chemical potential $\mu $ is
derived. Self-consistent approach is strictly necessary at $\mu \sim \Delta
_{nm}$ . Transition to the effective four-band model leading to the
temperature of superconducting transition is performed. Analytic and numeric
solutions have been performed for two mechanisms of superconducting
pairing -- non-phonon and phonon one. The account of the additional in
regards to Moscalenco model pairings of electrons from different bands or
their hybridization gives an option to obtain the high values of $T_{C}$
even at reduced density of charge carriers. This hybridization influences
greatly the form of the dependence of the quantity $T_{C}$ on the density of
charge carriers as well as the jump of the electron heat capacity at the
point $T=T_{C}$.

High and low values of the relative jump of electronic heat capacity at the
point $T=T_{C}$ $\left( \frac{\textstyle C_{S}-C_{N}}{\textstyle C_{N}}
>1,43\right. $ and $\,\left. \frac{\textstyle C_{S}-C_{N}}{\textstyle C_{N}}
<1,43\right) $ are obtained with varying the density of charge carriers.

These four effective energy bands in the considered system favors
essentially in experimental observation of the kink of the chemical
potential $\mu \left( T\right) $ at the point $T=T_{C}$ .

The mechanism of superconductivity (non-phonon and phonon) is shown to
affect differently with varying the density of charge carriers as well as
the dependence of $T_{C}$ and on this density.

Possible peculiarities in behavior of thermodynamic quantities in $MgB_{2}$
compound with reducing the density of charge carriers are predicted.

Influence of the overlapping energy bands on Fermi surface at $T=0$ and very
low densities of charge carriers $\left( \mu <0\right) $ on the
superconductivity in the system that is in the state of Bose condensation of
localized pairs is studied.

An application of the path integral method to the two-band model
is developed and on this basis, the process of transition from the Fermi to
the Bose pattern of elementary exitations at $T \not = 0$ in the presence of
a two-particll bound state in the system is demonstratied.The expression for
the temperature of Bose condensation $T_k$ is obtained and the contribution of
the residual boson interactions is estimated for systems with different
dimensions.

The overlappining of the energy bands on the fermi surface is favorable for
superconductivity and intensify some peculiarities of thermodinamic
properties at low carrier density. This halps to their experimental
confirmation.}

\end{quotation}

\begin{center}
\bf{1. INTRODUCTION}
\end{center}

A rich body of experimental and theoretical research material on oxide
ceramics has been accumulated since the discovery of high - temperature
superconductivity. However, the description of the physical properties of these
materials  remains one of the most difficult problems in low - temperature
physics today. This is because of the complexity of the objects of study: they
have a complex crystal structure, strong anisotropy, anomalies in the
electronic energy spectrum, a variable concentration of charge carriers, strong
electronic correlations, and so on. Models for the analysis of such systems
apparently should be based on the Hubbard model, which takes into account the
strong electronic correlations due to the Coulomb interaction of the electrons,
and should take into account the strong electron - phonon interaction. A
review of the different approaches to this problem and the approximations used
in them is given in Ref.1, for example. This theory contains dielectric and
magnetic phase transitions and the possible onset of superconductivity.
However, because of the great mathematical difficulties  it is hard to obtain
any meaningful physical results without making some substantial
simplifications. Moreover, at a certain carrier density a metallic states
arises in  the system, in which the electronic states are modified but not
destroyed by correlations. Consequently, there can be a transition to the
superconducting state, with the formation of Cooper pairs (the BCS scenario)
or local pairs (the Schafroth scenario). In this connection it is unquestionably
of interest to apply Fermi - liquid concepts to the study of the
superconducting properties of high - $T_c$ superconductors with allowance for
their peculiar features, such as overlapping of the energy bands at the Fermi
surface and the presence of various kinds of van Hove - Lifshitz singularities,
strong anisotropy, variable carrier density (including small values) etc., when
treating both phononic and nonphononic mechanisms of superconductivity.

Many theoretical papers have by now been published on various aspects of the
two - band model \cite{Moskalenko_2},\cite{Suhl} . The great
interest in this model and in its generalizations is due, firstly, to band
calculations \cite{Krakauer}, \cite{Herman} showing that in metal - oxide
ceramics the energy bands on the Fermi surface overlap (a similar situation
obtains, apparently, also in systems with heavy fermions \cite{Baranov}),
and secondly to the possibility of using the aforementioned model to describe
the properties of systems with two groups of electrons (e.g., layered
compounds).

The main development stages of the superconducting theory of the systems with
overlapping energy bands are presented in comment \cite{Palistrant_1}.

The consideration of the overlapping of the energy bands leads not only to the
quantitative differenece of results from the case of one - band superconductor,
but in some cases to the qualitatively new results. For example:

1) In two - band superconductors high temperature superconductivity is possible
not only in the case of attractive interaction between the electrons, but even
if the interaction between the electrons has repulsive character $(\lambda_{n
m} < 0, n, m = 1 - 2)$, but relation $\lambda_{11} \lambda_{22} - \lambda_{12}
\lambda_{21} < 0$ is fulfilled.

2) In impurity two-band superconductors, for example, Anderson theorem is
violated at $\Delta_1 \not = \Delta_2$, and appears the dependence of
thermodynamical quantities on concentrations of non - magnetic impurity due to
the interband scattering of electrons on imputiry atoms.

3) In two - band superconductors appears collective oscillations of excitontype
Leggett mode,caused by the fluctuations of phase of order parameters for
different bands. In the three - band systems, and also in the two - band with
lowered density of charge carrier, which reduces to the the three - band model,
such oscillatory modes can be two.

4) On the basis of the theory of superconductivity with overlapping energetical
bands one can explane a great number of experimental results in High -$T_c$
materials.
In particular, by using the two - band model and assuming moderate values of
the coupling constants one can obtain high $T_c$, two energy gaps $2\Delta_1
/T_c > 3,5$ and $2 \Delta_2 / T_c < 3,5$, large values of negative $d ln T_c /
d ln V$ ($V$ is the volume), a positiv curvature of the upper critical field
near the transition temperature, and others  \cite{Lee_1} - \cite{Volovik}.
Furthemore, in the two - band model it is possible to describe the
decrease of $T_c$ with increase of the oxygen disorder, as well as when copper
atoms are replaced by a nonmagnetic dopant (A1, Zn, etc)\cite{Moskalenko_5} -
\cite{Moskalenko_7}.

5). An important role in the determination of the thermodynamic and magnetic
properties of a two-band superconductor is played also by the location of the
Fermi level, which is changed by doping or by introduction of oxygen.
Having assumed a non-phonon pairing mechanism of superconductivity as
well as phonon mechanism in the multi-band systems with lowered densities of
charge carriers the account of the peculiarities mentioned above is very
crucial.Particular  interest attaches to the possibility of obtaining a
bell-shaped dependence of $T_c$ and of the heat-capacity discontinuity $(C_S -
C_N)$ at the point $T = T_c$ on the carrier density \cite{Palistrant_2}-
\cite{Palistrant_4}. In the three-band model with a nonphonon sureconductivity
mechanism it is possible to obtain the ''step'' which is observed for $Y_1 Ba_2
Cu_3 O_{7 - \delta}$ in the dependence of $T_c$ on the carrier density
\cite{Kalalb_1}, \cite{Palistrant_5}. An investigation of the properties of
high-temperature ceramics, based on allowance for energy -band overlap of the
energy bands and for electronic topological transitions, was reviewed in Ref
\cite{Moskalenko_8}.  This review contains the classical achievements of the
problem.On can find there references to experimental research results that can
be described by allowance for the singularitiesin the electron energy spectrum
of complex systems.

An increase of the number of energy bands on the Fermi surface increases the
overall electron-state density and leads to the onset of an additional
interband electron-electron interaction that contributes to the onset of
superconductivity. This interaction violates the universal BCS relations and
leads to a substantial dependence of a number of physical characteristics on
the properties of an anisotropic system \cite{Moskalenko_8}-\cite{Hirch}.

The discovery \cite{Nagamatsu} of superconductivity in $MgB_2$ with a high
transition temperature $T_c \sim 39 K$ has attracted considerable attention.
The theoretical investigations \cite{Bud'ko}-\cite{Choi_3} as well
experimental observations \cite{Buzea}, \cite{Ponomarev} have led to
the conclution that $MgB_2$ is multi-band superconductor with fonon-mediated
BCS superconductivity mechanism.

Therefore, the known classical results for the system with
overlapping energy bands  can be used as a basis to describe the
superconducting properties of $MgB_{2}$.

Nowadays, many authors  rediscover unfortunately well-known already
results, obtained early on the basis of Moskalenko \cite{Moskalenko_2}
two-band model.At the same time the method, taking into account the
overlapping ennergy bands as well as the anisotropy of energy gap, is developed
in a number of studies  (see, for exampl \cite{Mishonov_1} -
\cite{Mishonov_3}). This method gives a good option to incorporate the
theoretical results to the experimental data about the specific heat in
$MgB_{2}$ .

As mentioned above, an interesting feature of the two-band model is
independence of the superconducting-transition temperature of the sign of the
inter-and intraband interaction constants. This model can therefore be used for
the usual electron-phonon mechanism of superconductivity as well as for a
mechanism based on the repulsion between carriers. In all the references cited
above, the two-band model can be used to describe the properties of
superconductors for which the relation $\mu \gg T_c$ is satisfied ($\mu$ is the
chemical potential). This relation is satisfied in a number of cases also in
high- temperature compound. The existence of this relation between $\mu$ and
$T_c$ makes it possible to use in the calculations an approximation diagonal in
the band indices,\cite{Moskalenko_12},\cite{Moskalenko_13} which leads to
neglect of off-diagonal paramerers such as $\Delta_{12}$ and $\Delta_{21}$.

In systems with low carrier density, however, the relation $\mu \gg T_c$ does
not hold. It becomes therefore necessary to develop a superconductivity theory
for two - band systems, without constraints on the Fermi energy. This is the
task of the present paper. We consider simultaneously two possible
superconductivity mechanisms - phonon and electron. A characteristic feature of
systems with low carrier density is a substantial dependence of the chemical
potential $\mu$ on the order parameter in the superconducting phase. This
circumstances has been noted in many papers, and the feasibility of
expererimentally observing anomalies in the temperature dependence of the
chemical potential was first suggest in Ref.40. It was shown there, with the
BCS model as the example, the $\mu (T)$ curve has an experimentally observable
bend at the point $T = T_c$. We shall show below that in the two - band case
this effect is enhanced by the presence of four order parameters
$(\Delta_{nm}; m = 1,2)$ and is manifested at $\mu$ values easier to observe
in experiment.

In the articles referred to above, investigations were conducted on the basis
of the idea of Cooper pairing. In systems with small carrier concentrations,
such as semiconductors or metallic oxides, bound states
may arise following a decrease in the carrier concentration, and a transition
to a Bose condensate of local pairs with a final binding energy may occur (the
Schaffroth scenario \cite{Schaffroth}). The issue of such transitions
in one-band systems was discussed in a number of  articles \cite{Migake} -
\cite{Loktev}.

To realize the Schaffroth scenario, a bound two-particle state
\cite{Leggett_2} must exist in the system. The appearance of this state in
the presence of attractive interaction depends on the dimension of the system
\cite{Leggett_2} and is realized in the region of small carrier
concentrations. As is shown in \cite{Randeria}, \cite{Gordar} , the
change of sign  in the chemical potential  with a decrease in the carrier
concentration corresponds to a transition from the BCS to the Schaffroth
scenario. Condensation of local pairs occurs at concentrations of carriers for
which $\mu \leq 0$.

This review article is devoted to investigations of superconducting ordering
in systems with two characteristic features - a small concentration of charge
carriers and overlapping energy bands on the  Fermi surface. As mentioned
above, both of these features are characteristic of superconducting compounds.

Our review article is organized as follows. In Sec.2 we prezennt a theory of
the superconductivity of two-band system that is valid for any carrier density
and takes into account all possible pairing of electrons due to intraband and
interband interactions on the basis of the idea of Cooper pairing. The critical
temperature $T_c$, the chemikal potential $\mu$, the heat capasity  $(C_S -
C_N)$ at the point $T = T_c$ as function of carrier density are ilustrated.
Section 3 is devoted to a self-consistent discussion in the mean field
approximation of a system of equations in oreder parameters $\Delta_n$ and
$\mu$ at $T = 0$ and to revealing the influence of energy band overlapping on
these quantities and the carrier concentration at which the system experiences
a transition from Cooper pairing $(\mu > 0)$ to the Schaffroth $(\mu < 0)$
scenario. The equation for the binding energy $\varepsilon_b$ of a two-particle
state is also obtained and the relationship between $\varepsilon_b$ and $\mu$
is established. In Sec. 4 the path integral method as applied to the two-band
model is developed and, on this basis, the procedure for a transition from the
Fermi to Bose elementary excitations is given. The condensation temperature of
the Bose system $T_k$ is also determined. In the last section, the results of
the investigation are summarized.

\begin{center}
\bf{2. THERMODINAMICAL  PROPERTIES ON BASIS OF THE  COOPER PAIRS OF THE
CHARGE CARRIERS}.
\end{center}

{\bf 2. 1. System Hamiltonian and basic equations.}

The considered two-band system is described by the Hamiltonian \cite{Kochorbe_1}
$$
H = \sum_{n \vec k \sigma}\left[\varepsilon_n (\vec k) - \mu \right] a_{n \vec
k \sigma}^{+}\, a_{n \vec k \sigma} -
\frac{1}{V}\,\,\sum_{m_1...m_4,\vec k \vec k'}\,V_{m_1 m_2}^{m_3 m_4}\left(\vec
k - \vec {k'}; - \vec k \vec {k'}\right)
$$
\begin{equation}
a_{m_1 \vec k \uparrow} a_{m_2 - \vec k \downarrow} a_{m_3 - \vec k'\downarrow}
a_{m_4 \vec k'\uparrow}, \label{1} \end{equation} were $a_{n \vec k
\sigma}^{+}$ and $a_{n \vec k \sigma}$ are creation and annihilation
operators for a electron band $n$ with spin $\sigma$ and quasiwave vector {\bf
k}, $\mu$ is the chemical potential, and $V_{m_1 m_2}^{m_3 m_4}$ are the intra-
and interband interaction constants.  Expression (1) is a generalization of the
BCS - Bogolyubov model Hamiltonian to include the two-band case. Account is
taken here of all possible  methods of electron pairing within each band and of
electron pairing from different bands.  If $m_1 = m_2$ and $m_3 = m_4$, the
Hamiltonian (1) is equal to that of the Moskalenko model \cite{Moskalenko_2},
wich considers only intraband pairing and transitions of Cooper pair as a whole
from one band to another are considered; this model is widely used to describe
the properties of high-temperature superconductors \cite{Lee_1} - \cite{Hirch}
and magnezium deboride \cite{Jiu_1} - \cite{Choi_3}; \cite{Mishonov_1} -
\cite{Mishonov_3}.Examination of the more general Hamiltonian (1) uncovers
additional possible onsets of superconductivity (on account of single-particle
hybridization and of all interband-interaction constants) and makes passible a
description of the properties of a system with a low density of states $(\mu
\sim T_c)$.Applying the Green's- function method \cite{Abricosov_1}  to the
Hamiltonian (1) we obtain the set of equations for the order parameters
$\Delta_{np}$ \cite{Kochorbe_1}.

\begin{eqnarray}
\Delta_{np}&=& \frac{1}{4V}\,\sum_{\vec klr, r\not= l}\,V_{np}^{ll} \Delta_{ll}
\Bigg[\left(\frac{\xi_{l}^{2} -\xi_{r}^{2} + 2\Delta_{12} \Delta_{21}(1 +
1/z_{(12)}}{d} + 1 \right)\nonumber\\
&\times& th \frac{\frac{\beta E_1}{2}}{E_1} -  \left(\frac{\xi_{l}^{2}
- \xi_{r}^{2} + 2\Delta_{12} \Delta_{21}(1 + 1/z_{(12)}}{d} - 1 \right)
th \frac{\frac{\beta E_2}{2}}{E_2}\Bigg]+
\nonumber
\end{eqnarray}
\begin{equation}
+ \frac{1}{4V}\,\sum_{\vec k}\left(V_{np}^{12}\Delta_{12}\frac{z_{(13)}}
{z_{(14)}}+ V_{np}^{21}\Delta_{21}\,\frac{z_{(14)}}{z_{(13)}}\right) \Gamma,
\label{2}
\end{equation}
where
\begin{equation}
E_{1,2} = \sqrt{\frac{a \pm d}{2}},\,\,a = \xi_{1}^{2} +\xi_{2}^{2} + 2
\Delta_{12} \Delta_{21},
\label{3}
\end{equation}
$$
d^2 = (\xi_{1}^{2} -\xi_{2}^{2}) + 4 \Delta_{12}\Delta_{21}\left[
(\bar \varepsilon_1 - \bar \varepsilon_2)^2 + (\Delta_{11} + \Delta_{22})^2
\right],
$$
$$
\Gamma = \left[\frac{(\bar \varepsilon_1 - \bar \varepsilon_2)^2 + (\Delta_{11}
+ \Delta_{22})^2}{d} + 1 \right] \frac{th\frac{\beta E_1}{2}}{E_1} -
$$
\begin{equation}
- \left[\frac{(\bar \varepsilon_1 - \bar \varepsilon_2)^2 +
(\Delta_{11} + \Delta_{22})^2}{d} - 1 \right] \frac{th\frac{\beta
E_2}{2}}{E_2};
\label{4}
\end{equation}
$$
\xi_{n}^{2}=\xi_{n}^{2} \vec k = \bar \varepsilon_{n}^{2} +
\Delta_{nn}^{2},\,\,\bar \varepsilon_n =\varepsilon_n - \mu_{n},\,(n\,;
p\,;l\,;r = 1,\, 2) ,
$$
\begin{equation}
z_{(12)} = \Delta_{11}/\Delta_{22},\,\,z_{(13)} = \Delta_{11}/\Delta_{12},\,\,
z_{14} = \Delta_{11}/\Delta_{21}.
\label{5}
\end{equation}
The system (2) determines the order parameters $\Delta_{11}$ and $\Delta_{22}$
of an ordinary two-band superconductor (and can be simplified in this case by
putting $\Delta_{12} = \Delta_{21} = 0$) as well as of a superconductor with
low carrier density $\mu \sim T_c$. We supplement the system (2) with the
expression \cite{Kochorbe_1}.
\begin{equation}
N_0 = \sum_{knm,n\not=m} \left[1 - \left(\frac{\bar \varepsilon_n +  \bar
\varepsilon_m}{2} + \frac{\bar \varepsilon_n - \bar \varepsilon_m}{2}
\frac{\xi_{1}^{2} - \xi_{2}^{2}}{d}\right) \frac{th\frac{\beta
E_n}{2}}{E_n} \right].
\label{6}
\end{equation}

The self consistent set of Eqs. (2) and (6) determines the order parameters
$\Delta_{np}$ and the chemical potential $\mu$ for a specified temperature
$T$ and a carrier density $N_0$. A characteristic feature of the ground state
in a system with low carrier density is a substantial change of the position
of the Fermi level following the onset of the superconducting gap. The order
parameters $\Delta_{np}$ become of the same order as the chemical potential
$\mu\, (\mu \sim \Delta_{np})$. This leads to an anomalous behavior of the
chemical potential as a function of temperature. In particular, in the single-
band BCS model \cite{Van} and in the Hubbard model \cite{Robashkiewicz}
the chemical potential has in rarefied systems a kink at the point $T = T_c$.
As first noted by Van der Marel \cite{Van}, this kink is observable in
experiment and consequently its observation can help explain the
superconductivity mechanism. In Ref.\cite{Van} has demonstrated the
possibility of a kink at $\mu \leq 2\, me V$, which is the lower limit of
present-day accuracy \cite{Cardona}. Since the overlap of the energy bands
plays a major role in the explanation of the properties of high-temperature
supercondutors, it is of interest to investigate the anomalous behavior of the
chemical potential in a two-band model with low carrier density, and the
possible onset of a kink at values of $\mu$ more conducive to experimental
observation. We represent the order parameter $\Delta_{nm}$ near the
superconducting transition temperature $T \sim T_c$ in the form
\begin{equation}
\Delta_{nm} = c_{nm}(\beta - \beta_c)^{1/2} + c_{nm}^{1}
(\beta - \beta_c)^ {3/2} + ...\,\, .
\label{7}
\end{equation}
The expression  for the chemical potential $\mu$ near $T_c$ is now
\begin{equation}
\mu(T) = \mu_0 (T) - R_0 \Delta_{11}^{2},
\label{8}
\end{equation}
were $\mu_0(T)$ is the chemical potential of the normal phase,
and $R_0$ is determined from the constancy of the carrier density in the
superconducting and normal phases.  Substituting (7) in (8) we readily obtain a
jumplike change of $d \mu / d T$.  Contributions to this change are made by all
the order parameters $\Delta_{nm}$, so that the results can differ from those
for the case of one band \cite{Van}.\\
\\
{\bf 2. 2. Self-consistent system of equations for T = Tc}

To investigate the properties of a two-band system near the superconducting -
transition temperature we expand in Eqs.(2) and (6)in terms of the small
parameters $\Delta_{nm}\, \Delta_{mn}$ with account taken of expansions (7)
and (8).

We introduce the dispersion law of the $n$th band
\begin{equation}
\varepsilon_n = - \zeta_n - \frac{k_{x}^{2} + k_{y}^{2}}{2 m_n},
\label{9}
\end{equation}
and change in these equations from summation over $\vec k$ to integration over
the energy in accordance with the dispersion law (9)$\left(\bar \varepsilon_{n
0} = \varepsilon_n - \mu_{0 n}, \,\mu_{0 n} = \mu_0 - A_n \right)$:

\begin{equation}
\frac{1}{V}\,\sum_{\vec k}\,\Phi (\varepsilon_n - \mu_{0 n}) = 2
N_n\,\int_{ - D_n}^{D_{cn}} \,dx \Phi(- x),
\label{10}
\end{equation}
where $N_n = m_n k_{zn}^{0} /2 (2\pi)^2$ is the density of the electron states
in the $n$th band, $A_n$ is a quantity that renormalizes the chemical potential
in the self-consistent-field approximation \cite{Kochorbe_1}. The integration
limits are chosen to be able to consider simultaneously two possible
superconductivity mechanisms: the values of $D_n = \eta_{0 n} - \zeta_n \leq
\omega_{D_n}$ and $D_{cn} = \omega_{D_n}$ ($\omega_{D_n}$ is the phonon cutoff
frequency in the $n$th band) correspond to the phonon mechanism of
superconductivity, and the quantities $D_n = \eta_{0n} - \zeta_n$ and $D_{cn}
=\zeta_{cn} - \eta_{0n}$ ($\zeta_{cn}$ is a cutoff energy of the order of that
of the electron;$\eta_n = - \mu_{0n}$) corresponds to the hole mechanism.

Integrating next over the energy and equating the coefficients of equal powers
in the difference ($\beta - \beta_c$), we obtain for $c_{nm}$ and
$c_{nm}^{(1)}$ the set of equations \cite{Kochorbe_1}.  In the "pseudoband"
representation these equations take formally the form of the four - band model
\cite{Moskalenko_2}, \cite{Moskalenko_9}.
\begin{equation}
c_{(n)} = - \sum_{(m) = 1}^{4}\,N_{(m)} J_m
U_{(n)(m)}c_{(m)},
\label{11}
\end{equation}
\begin{equation}
c_{(n)}^{1} = - \sum_{(m) = 1}^{4}\,N_{m} J_m
U_{(n)(m)}c_{(m)}^{(1)} - \sum_{(m) = 1}^{4}\,\,\Phi_{(m)}\,U_{(n)(m)},
\label{12}
\end{equation}
where
\begin{equation}
\Phi_{(m)} = \left(-\frac{\Theta_{(m)}}{\beta} + (\beta c_{(m)})^2 \bar F_{m})
\right) N_{(m)}.
\label{13}
\end{equation}
At $m = 1,2$ we have
$$
J_m = J^0 (- \beta D_{cm}) - J^0 (\beta D_m),\,J^0(x) = \int
\limits_{0}^{x}\,\,\frac {th\,y/2}{y}dy
$$
\begin{equation}
\Theta_m = th \frac{\beta\,D_{cm}}{2} - th \frac{\beta\,D_{m}}{2}\,,n = 1 - 4.
\label{14}
\end{equation}

The functions $N_{(m)}\,\bar F_{(m)}$ at m = 1 - 4 and $N_{(m)}\,J_m,\,\,\,
N_{(m)}\,\Theta_m$ at $m = 3,4$ are complicated functions, contained the
summations on zon indexis \cite{Kochorbe_1}. We introduce quasiband indices
in accord with the rule
\begin{equation}
11 \rightarrow (1),\,\,\,22  \rightarrow (2),\,\,\,12 \rightarrow (3),\,\,\,21
\rightarrow (4),
\label{15}
\end{equation}
and also the symbols:
$$
V_{nm}^{pr} \rightarrow V_{nm\,\,pr} \rightarrow U_{(n')(m')};\,\, n^{/},
m^{/} = 1 - 4.
$$
We omit hereafter the parentheses of the pseudoband  subscripts.  It is
convenient to rewrite Eqs.(11) and (12) in the matrix form
\begin{equation}
D_0 c = 0
\label{16}
\end{equation}
\begin{equation}
D_0 c^{(1)} = - \Phi,
\label{17}
\end{equation}
were $c,\,c^{(1)}$, and $\Phi$ are single - columm matrices in the indices 1
to 4, and
\begin{equation}
D_0 =
\left( \begin{array}{cccc}
1 + N_1J_1U_{11} & N_2J_2U_{12} & N_3J_3U_{13} & N_4J_4U_{14}\\
N_1J_1U_{12} & 1 + N_2J_2U_{22} & N_3J_3U_{23} & N_4J_4U_{24}\\
N_1J_1U_{31} &  N_2J_2U_{32} & 1 + N_3J_3U_{33} & N_4J_4U_{34}\\
N_1J_1U_{41} &  N_2J_2U_{42} &  N_3J_3U_{43} & 1 + N_4J_4U_{44}
\end{array} \right).
\label{18}
\end{equation}
From the condition that the system (16) have a solution, we obtain an equation
for the critical temperature $T_c$:
\begin{equation}
||D_0|| = 0
\label{19}
\end{equation}
where $||...||$ designates the determinant of a matrix. It follows from the
system (17) that
\begin{equation}
c_{1}^{(1)} = ||D_1||/||D_0||.
\label{20}
\end{equation}
Since $||D_0|| = 0$, we get
\begin{equation}
||D_1|| = 0,
\label{21}
\end{equation}
where $D_1$ is a $4 \times 4$ matrix that differs from $D_0$ in that the first
column is replaced by the elements of the matrix $\Phi$.
We have on the basis of (20) the expression for $c_{1}^{2}$
\begin{equation}
c_{1}^{2} = \frac{1}{\beta_{c}^{3}}\,\frac{\sum_{n = 1} N_n \theta_n f_n
/ z_{1n}^{2}}{\sum_{n = 1}^{4} N_n \overline{F}_n f_n / z_{1n}^{4}} = \frac
{\bar c_{1}^{2}}{\beta_{c}^{3}}.
\label{22}
\end{equation}

Assuming the particle number $N_0$ to be fixed and using the expansion (7) for
the chemical potential near $T_c$ we obtain
\begin{equation}
\mu (T) = \mu_0 (T) - \frac{R_0}{\beta}\,{\bar c_{1}^{2}}(T_c - T),
\label{23}
\end{equation}
and $\mu_0$ is determined from the equation
\begin{equation}
N_0 = 2 \sum_{l} \left[D_{cl} + D_l - \frac{2}{\beta}ln \frac{ch\frac{\beta
D_{cl}}{2}}{ch\frac{\beta D_l}{2}}\right].
\label{24}
\end{equation}
The determinations of the functions $f_n / z_{1n}$ and $R_0$ see in
\cite{Kochorbe_1}.

It is possible to calculate the dependence of the superconducting temperature
$T_c$ on the carrier density $N_0$ for all values of the interaction constants
$\lambda_{nm} = N_m U_{nm}\, (n,m = 1 - 4)$, and also the temperature dependence
of the chemical potential $\mu(T)$ on basis of (18) - (24).

Figure 1 shows the dependence of the superconducting temperature $T_c$ on the
carrier density $N_0 / 2 N_1$ for different degrees of hybridization in
the casse of electron mechanizm of superconductivity:  a) weak:  $\lambda_{11} =
\lambda_{22} = 0.2,\,\lambda_{12} = \lambda_{21} = \lambda_{33} = \lambda_{44}
= 0.01,\,\lambda_{34} = \lambda_{43} = 0.105$, the remaining ones:$\lambda_{nm}
= 0.001\,(n, m = 1 - 4)$; b) strong $\lambda_{11} = \lambda_{22} = 0.2,
\lambda_{12} = \lambda_{21} = \lambda_{33}= \lambda_{44} = 0.1,\, \lambda_{34}
= \lambda_{43} = 0,15 ,$ the remaining ones $\lambda_{nm} = 0.01\,(n, m = 1 -
4)$.

Calculations are perfomed for the following values of the parameter
$$
\zeta_1 = 0\,eV,\,\,\zeta_2 = 0.03\,eV,\,\,\zeta_{c1} = 0.05\,eV,\,\,
\zeta_{c_2} = 0.08\,eV.
$$

It follows from Fig.1 that one can obtaina bell-shaped dependence of $T_c$ on
$N_0$ (curves 1-3) as well as the weak dependence $T_c (N_0)$ curves(1 - 3).
Inclusion of strong hybridization raise $T_c$.  The character of the
dependence of the transition temperature $T_c$ on the carrier density $N_0$ is
strongly influenced by the relation between the electronic - state densities of
the different bands.  Lowering the ratio $N_2 / N_1$ slows down the growth of
$T_c$ with increase of $N_0$ (for case b) and accelerates the decrease for both
cases $(a,b)$

In the case of weak hybridization (curves $1 - 3$) the $T_c\,(N_0)$ plot
acquires two maxima. The degree to which then become pronounced, given the
parameters $\lambda_{mn}$, is determined by the electron state-density ratio
$N_2 / N_1$. The presence of weakly pronounced maxima (curve 1) corresponds to
the case $N_1 = N_2$, and more strongly pronounced maxima (curves 2 and 3)
appear at $N_1 \not = N_2$ and are determined by the anisotropy of the system
(by the difference between the bands). Each of these maxima is connected with
the occupation of the corresponding band. In the absence of interband
interaction (energy - band overlap) the plot would consist of two nonoverlapping
curves. The onset of interband interactions produces simultaneos
superconductivity in both bands, with a single superconducting temperature
determined by all the interaction constants $\lambda_{nm}$. With increase of
the interband-interaction constants $(\lambda_{34},\lambda_{43} \sim
\lambda_{11},\lambda_{22})$ the contribution due to overlap of the two bands
begins to predominate over the individual contribution of each band, so that
the $T_c (N_0)$ plot is a single bell - shaped curve (4,5).

Figure 2 shows the temperature dependence of $\eta = - \mu$ for a non - phonon
superconductivity mechanism at various carrier densities $N_0$ and with weak
hybridization. We see that this plot has at $T = T_c$ (curves 1 - 3) a kink
that becomes less peaked with increase of the carrier density and vanishes at
$\eta \sim 8$ me V (curve 4). The behavior of $\eta (T)$ under strong
hybridization is similar. The anomaly of the temperature dependence of the
chemical potential $\eta = - \mu$ at the point $T = T_c$ is due to the
appearance, on the Fermi surface, of a superconducting gap that does not differ
excessively from the chemical potential. This gap influences substantially the
chemical potential $\mu$ at $T < T_c$, since the values of $\eta$ and
$\Delta_{n m}$ are self-consistently determined from the set of equations (2)
and (6).

Allowance for the energy-band overlap leads to a kink on the temperature
dependence of the chemical potential at $T = T_c$ for the sufficiently high
values $\eta \leq 8 meV$ me ($\eta \leq 2 meV$ in the single band case
\cite{Van}, which undoubtedly facilitates experimental verification of this
effect).
\\\\
{\bf 2. 3. Critical temperatur $T_c$ and ratio $2\Delta_n / T_c$ \\
\,\,\,\,\,\,in the limiting case $\Delta_{12} = \Delta_{21} = 0$.}

In the limiting case  $\Delta_{12} = \Delta_{21}\,\,\,(N_3 = N_4 = 0)$ the
problem is simplify essentialy . Such model corresponds to  \cite{Moskalenko_2},
\cite{Moskalenko_9}  the equation for $T_c$ has the following form
\cite{Palistrant_2}:

\begin{equation}
1 + \lambda_{11} J_1 + \lambda_{22}J_2 + a J_1\,J_2 = 0,
\label{25}
\end{equation}
\begin{equation}
\lambda_{nm} = V_{nm} N_m,\,\,a = \lambda_{11} \lambda_{22} - \lambda_{12}
\lambda_{21}.
\label{26}
\end{equation}
Besides, the $\eta$ in the case of nonfonon mechanism of superconductivity is
defined from
\begin{equation}
N_0 = 2\,\sum_n N_n \left[\zeta_{cn} -
\zeta_n - |\zeta_{cn} - \eta| + |\zeta_{n} - \eta| - 2 T_c ln \frac{1 +
exp(-\beta_c |\zeta_{cn} - \eta|)}{1 + exp (- \beta_c|\zeta_n - \eta|)}\right].
\label{27}
\end{equation}

From the relation (22) we have
\begin{equation}
c_{1}^{2} =  \frac{1}{\beta_{c}^{3}}\,\frac{N_1\,\theta_1 \,+
N_2\,\theta_2 / z^2}{N_1 F_1 + N_2 F_2 / z^4} ,
\label{28}
\end{equation}
where you $J_m$ and $\Theta_m$ are determined by (14), and
\begin{equation}
F_m = -\frac{1}{4} \int \limits_{-\beta_c  D_m}^{-
\beta_c D_{cm}}\, dx \, \frac{shx - x}{x^3 ch^2 x/2}\,\,,z = \Delta_1/\Delta_2.
\label{29}
\end{equation}

For the sake of definiteness we choose $\zeta_1 = 0 < \zeta_2 <
\zeta_{c1} < \zeta_{c2}$ and consider the equation for the
superconducting transition temperature (25). On the basis of this equation one
can obtain the analytic expressions for the transition temperature $T_c$ if
the following conditions are satisfied: $|\eta - \zeta_n|/ T_c \gg 1$ and
$|\zeta_{cn} - \eta|/ T_c \gg 1$
So, for definite values of $\eta$ we obtain the following expressions:\\
(1) For $0 < \eta < \zeta_2$ :
\begin{equation}
T_c = \frac{2 \gamma}{\pi}\,\sqrt{\eta(\zeta_{c1} - \eta)}
exp\,\left[- \frac{1}{2}  \frac{1 + \lambda_{22} ln ((\zeta_2 -
\eta)/(\zeta_{c2} - \eta))}{\lambda_{11} + a\,ln((\zeta_2 -
\eta)/(\zeta_{c2} - \eta))}\right] ,
\label{30}
\end{equation}

$N_0 = 4 N_1 \eta$.

(2) For $\eta = \zeta_2$:
$$
T_c = \frac{2 \gamma}{\pi}\,\eta^{1/4}(\zeta_{c_1} -
\eta)^{1/4}\sqrt{\zeta_{c_2} - \eta}\,
$$
\begin{equation}
exp\,\left[- \frac{2 \lambda_{11}  + \lambda_{22}}{4a} \pm \frac{1}{4a}
\sqrt{\left(a\,ln \frac{\eta (\zeta_{c_1} - \eta)}{(\zeta_{c_2} -
\eta)^2} + 2 \lambda_{11} - \lambda_{22}\right)^2 + 8\lambda_{12}
\lambda_{21}}\,\,\right],
\label{31}
\end{equation}

$N_0 = 4N_1 \eta$.

(3) For $\zeta_{c2} < \eta < \zeta_{c1}$:

$$
T_c = \frac{2 \gamma}{\pi}\left[\eta (\zeta_{c_1} - \eta)(\zeta_{c_2}
- \eta)(\eta - \zeta_2)\right]^{1/4}
$$
\begin{equation}
exp\,\biggl\{- \frac{\lambda_{11} + \lambda_{22}}{4 a} \pm \frac{1}{4 a}
\sqrt{ a\,ln\, \frac{\eta (\zeta{c_1} - \eta)}{(\zeta_{c_2} -
\eta)(\eta - \zeta_2)} + (\lambda_{11} -\lambda_{22})^2 + 4 \lambda_{12}
\,\lambda_{21}} \biggr\},
\label{32}
\end{equation}

$N_0 = 4 N_1\,\eta_1 + 4 N_2 (\eta - \zeta_2)$.

(4) For $\eta = \zeta_{c1}$

$$
T_c = \frac{2\gamma}{\pi} \sqrt{\eta}\,(\zeta_{c2} - \eta)^{1/4} (\eta -
\zeta_2)^{1/4}
$$
\begin{equation}
 exp\,\left[ -\frac{\lambda_{11} + 2 \lambda_{22}}{4a} \pm \frac{1}{4 a}
\sqrt{\left(a\,ln\,\frac{\eta^2}{(\zeta_{c_2} - \eta) (\eta -
\zeta_2)} + \lambda_{11} - 2 \lambda_{22} \right) +
8\,\lambda_{12} \lambda_{21}}\right],
\label{33}
\end{equation}

$N_0 = 2 N_1 (\eta + \zeta_{c_1}) + 4N_2(\eta - \zeta_2)$.

(5) For $\zeta_{c_1} < \eta < \zeta_{c_2}$

\begin{equation}
T_c = \frac{2 \gamma}{\pi} \sqrt{(\zeta_{c_2} - \eta)(\eta -
\zeta_2)} \,exp \left[ - \frac{1}{2}\,\frac{1 + \lambda_{11}\, ln ((\eta
- \zeta_{c1}) / \eta)}{\lambda_{22}+ a\,ln ((\eta - \zeta_{c1})
/\eta)}\right],
\label{34}
\end{equation}

$N_0 = 4 N_1\,\zeta_{c_1} + 4 N_2 (\eta - \zeta_2)$.

The expressions $(31) \sim (33)$ contain two solutions wich correspond to the
sign $"\pm"$ in the brackets. A negative value of the quantity in the exponent
is a necessary condution for the solution's selection. In the case where both
solutions satisfy the above necessary condition, the solution giving the
greater $T_c$ must be selected. Note that we will obtain the results of Ref.
\cite{Moskalenko_2} provided $V_{11} = V_{22}  = 0$ and $\zeta_{c_1} =
\zeta_{c_2}$.

The analysis of the above $T_c$ expressions permits us to make the conclusion
that high $T_c$ values can be achived both for $\lambda_{nm} > 0$ (carrier
attraction) and $\lambda_{nm} < 0$ (repulsion). In the latter case the
condition $ a = \lambda_{11} \lambda_{22} - \lambda_{12} \lambda_{21}< 0$,
imposing restrictions on the constants $\lambda_{nm}$, must be satisfied.
The dependence of $T_c$ on $\eta$ is demonstrated in fig. 3 at definite values
of the parameters $\lambda_{nm}$. Here and hereafter in computations we choose
the following values of the parameters: $\zeta_1 = 0 \,e V,\,
\zeta_2 = 0.11 \,e V,\,\zeta_{c_1} = 0.2 \,e V$\,and
$\zeta_{c_2} = 0.3 \,e V\, $. Curves 1 and 2 correspond to the case
$\lambda_{nm} > 0$ and to the case $\lambda_{nm} < 0$, respectively. As follows
from this figure, high values of $T_c$ can be achieved as the chemical
potential changes. The $T_c$ dependence on $\bar N_0 = N_0 / 4 N_1$ can be
presented easily. In this case the ratio $N_2 / N_1$ must be given. Then the
$T_c$ dependence will be defined by this ratio. The $\eta$ dependence  on $\bar
N_0$ at different values of $N_2 / N_1$ is shown in fig.4. As follows from this
figure, the rate of growth of the quantity $\eta$ is defined by the ratio
$N_2 / N_1$ as $N_0$ increases.
The critical temperature $T_c$ as well as the order paramiters $\Delta_m$ as
function of $\eta$ in this limit case is studied detaly in
\cite{Palistrant_2}.

This investigations permit us to define the ratios $2 \Delta_1 / T_c$ and
$2 \Delta_2 / T_c$ as function of $\eta$ (see fig.5). The curves 1 and 2
correspond to the ratios $2 | \Delta_1 (0) | / T_c$ and
$2 | \Delta_2 (0) | / T_c$ respectively for $\lambda_{nm} > 0$, and the curves $1^{/},2^{/}$
correspond to the ones for $\lambda_{nm} < 0$. As is seen from
this figure the behavior of the quantities $2 | \Delta_1 (0) | / T_c$ and
$2 | \Delta_2 (0) | / T_c$ as functions of $\eta$ depends essentially on the
values of the parameters $\lambda_{nm}$.When $\lambda_{nm} > 0 $  we have a
step decrease of $2 |\Delta_1(0)|/T_c$ and a smooth growth of $2 |\Delta_2
(0)| / T_c$ as $\eta$ increases. When $\lambda_{nm} < 0 $ the ratio $2 |
\Delta_1 (0) | / T_c$ essentially increases and can achieve values near 7.5
and the ratio $2 | \Delta_2 (0) | / T_c$ slowly decreases, achieving the value
3.5 as $\eta$ increases. So the values of $2 | \Delta_1 (0) | / T_c$ and $2 |
\Delta_2 (0) | / T_c$ depend on $\eta$ and can be essentially different from
the BCS theory ones where $2 | \Delta (0) | / T_c$ = 3.5  and from the two -
band theory with phonon superconductivity mechanism \cite{Moskalenko_2},
\cite{Moskalenko_9}, where these ratios are independent of the chemical
potential (carrier concentration).  Note that for $T_c$, as for $| \Delta_n
(0)|$, the ratio $2 | \Delta_n (0)| / T_c$ can be shown in dependence on the
carrier concentration, by using relations between $\eta$ and $\bar N_0$ (see
fig.4).
\\\\
{\bf 2.4. Heat - capacity jump at the point T = Tc.}

In the section 2.2 we changed over to the pseudo-band representation, which
allowed us to write down Eqs. (11) and (12), as  it were, for a four-band
model.

It can also be shown that the difference between the free energies in the
superconducting and normal phases in the pseudoband representation generalizes
the corresponding expression of the Moskalenko two-band
model \cite{Moskalenko_2}, \cite{Moskalenko_9}.  We obtain

\begin{equation}
\frac{\Psi_S -  \Psi_N}{V} = \sum_{nmp}\,\int_{0}^{\Delta_p}\,\Delta_n \Delta_m
\frac{(\delta U^{-1} )_{nm}}{\delta \Delta_p} \delta \Delta_p,
\label{35}
\end{equation}

where $n, m$ and $p$ are the pseudoband nambers $(n, m, p = 1 - 4)$, and $U^{-
1}$ is the inverse of the interaction matrix $U$ in (11). We expand, in the
pseudoband formalism, the set of Eqs. (2) for the order parameters $\Delta_n$
in powers of the small quantity $(\beta \Delta_n)^2$ in the vicinity of the
critical temperature $T_c$:
\begin{equation}
\Delta_m = - \sum_n N_n\,U_{nm} (J_n + (\beta \Delta_n)^2 \overline F_n + ...),
\label{36}
\end{equation}
where $J_n$ and  $\overline F_n$ are defined in [49]. Using the calculation method of
Refs.\cite{Moskalenko_9} and \cite{Palistrant_2} for Eqs.(35) and (36),
we obtain for the heat - capacity jump at the point $T = T_c$
$$
\frac{C_S -  C_N}{V} = - T \frac{\partial^2}{\partial\, T^2} \frac{\Psi_S -
\Psi_N}{V} = \beta_{c}^{5}\sum_{n = 1}^4 N_n\,\overline F_n\,c_{n}^{4} =
$$
\begin{equation}
= T_c   \left[\frac  {\Sigma_{n = 1}^{4}\,N_n \theta_n f_n / z_{1n}^{2}}{
\Sigma_{n = 1}^{4}\,N_n \overline F_n f_n / z_{1n}^{4}}\right]^2\,\sum_{n = 1}^
{4}\,\,\frac{N_n\,\overline F_n}{z_{1n}^{4}}.
\label{37}
\end{equation}

The equation for the electronic heat capacity in the normal phase is
\cite{Palistrant_2}
\begin{equation}
C_N = 4 T\,\sum_{n = 1}^{2}\,N_n \phi_n (\eta),
\label{38}
\end{equation}
where
\begin{equation}
\phi_n (\eta) = \int_{- \beta (\eta - \zeta_{cn})}^{\infty}\,\,\frac{x^2 d x}
{(1 + e^x)(1 + e^{- x})}.
\label{39}
\end{equation}
In accordance with Eqs. (37) and (38) we have for the relative heat - capacity
jump
\begin{equation}
\frac{C_S - C_N}{C_N} = \left[\frac  {\Sigma_{n = 1}^{4}\,N_{n} \theta_
{n} f_{n} / z_{1n}^{2}}{ \Sigma_{n = 1}^{4}\,N_{n} \overline F_{n}
f_{n}/ z_{1n}^{4}}\right]^2\, \frac  {\Sigma_{n =
1}^{4}\,N_{n} \overline F_n / z_{1n}^{4}}{4 \Sigma_{n = 1}^{2} N_n \,\phi_n
(\eta)}.
\label{40}
\end{equation}
Figures 6 shows the dependences of the absolute and relative jump of the
electron heat capacity at the point  $T = T_c$ on the carrier density $N_0$,
obtained by numerical methods using the equations given above. The numbers of
the curves in these figures are the same as in Fig.1.

As seen from Fig. 6a, the dependence of $(C_S - C_N)_{T = T_c}$ on $N_0$ has a
maximum. At the same time, this dependence does not duplicate the behavior
of $T_c (N_0)$. This circumstance indicates that a substantial contribution
to the dependence of $(C_S - C_N)_{T = T_c}$ on $N_0$ is made not only by
$T_c$ but also by the complicated function in the right-hand side of (37).
Analysis of the curves of Fig.6 shows that the charater of the plot of $(C_S -
C_N)_{T = T_c}$ versus $N_0$ is determined by the type of hybridization (strong
or weak) and by the ratio $N_2 / N_1$ of the electron - state densities. It is
possible for $T_c$ and $(C_S - C_N)_{T = T_c}$ to have maxima at one and the
same value of $N_0$ (curves 4 and 5). This situation is observed in
experiment, for example in $La_{2 - x}Sr_x Cu_2 O_4 $ \cite{Hongshum}.

The possibility of obtaining small $(C_S - C_N) / C_N < 1.43$, as well as large
$(C_S - C_N) / C_N >  1.43$ values of the relative jump of the electron
specific heat is demonstrated by Fig. 6b. This picture is observed in high -
temperature ceramics \cite{Junod} -\cite{Akis_2}. The complicated dependence
of the relative electron-heat-capacity jump, shown in Fig. 6a, is determined by
the competition between the behavior of the diference $C_S - C_N$ shown in
Fig.6a as a function of $N_0$ and the quantity $C_N$ which  increases as $N_0$
increases.

We have considired in the present study quasi-two-dimensional systems with a
simple dispersion law (9). This approach is dictated, in particular, by the
lower dimensionality of a number of high-temperature ceramics. Since, however,
the electron-state densities $N_n (\varepsilon) , n = 1, 2$ have no
singularities for the dispersion law (9), we obtain the very same equations
also for a three-dimensional system. Only the values of $N_n$ will differ. Just
as in the case of single-band superconductors, \cite{Van} in our case
the ratio $C_S-C_N > 1.43$ is goverened to a considerable degree by the small
$C_N$ in the considered range of $N_0$ compared with the case of ordinary
metals  (or by the faster increase of $C_S-C_n$ with increase of $N_0$ compared
with the increase of $C_S-C_N$ with increase of $N_0$ compared with the
increase of $C_N$).

In the case $N_{3} = N_{4} = 0$ (this limit corresponds to
ujualy two-band model \cite{Moskalenko_2}), we  presents the  specific heat jump
in the form Ref. (\cite{Palistrant_2}):
\begin{equation}
\frac{C_S - C_N}{V} = T_c
\frac{\left[ N_1\theta_1 (\eta) + N_2 \theta_2 (\eta) / z^2 \right]^2}{N_1
F_1(\eta) + N_2 F_2 (\eta) / z^4},
\label{41}
\end{equation}
\begin{equation}
\frac{C_S - C_N}{C_N} = T_c \frac{\left[ N_1\theta_1 (\eta) + N_2 \theta_2
(\eta) / z^2 \right]^2}{4 \left[N_1 F_1(\eta) + N_2 F_2 (\eta) / z^4\right]
\left[N_1 \varphi_1 (\eta) + N_2 \varphi_2 (\eta)\right]}.
\label{42}
\end{equation}
The presence of the functions $\theta_n (\eta)$ and $F_n (\eta)$ distinguishes
this expression from the two-band theory with phonon superconductivity. In the
above approximation $\eta >> T_c$, values of these functions at the different
values of $\eta$ are given in table 1 Ref.\cite{Palistrant_2}. So the specific
heat jump depends on $\eta$ (on $N_0$ through the $T_c$ dependence on this
quantity and by the suplementary function in the right hand part of eq. (41).
By using table 1  \cite{Palistrant_2}  and the above values of $T_c$ the
specific heat jump dependence $(C_S - C_N))$ on $N_0$ can be easily presented
on the basis of eq. (41). This dependence is shown in fig.7 at different values
of the ratio $N_2 / N_1$. The curves with non-primed numbers correspond to
$\lambda_{nm} > 0$ and the ones with primed nambers correspond to $\lambda_{nm}
< 0$. We obtain the result that the quantity $(C_S - C_N) / V N_1$ increas as
the carrier concentration increases.  The maximum of this quantity is achieved
at the values of $N_0$ wich corespond to the superconducting transition
temperature. Then this quantity decreases as $N_0$ increases for all values of
$N_0 / N_1$.It is interesting to note also that the same  dependence of the
absolute specific heat jump at the point $T = T_c$ on $S_r$ concentration has
been observed in $La_{2-x} S_{rx} Cu_2 O_4$ \cite{Hongshum}. By using table 1
of Ref.  \cite{Palistrant_2} we present expressions for the relative jump
of electron spesific heat in some points.\\
1) For $0 < \eta < \zeta_2$:
\begin{equation}
\frac{C_S - C_N}{C_N} = \frac{12}{7 \zeta(3)} = 1,43.
\label{43}
\end{equation}
2) For $\eta = \zeta_2$:
\begin{equation}
\frac{C_S - C_N}{C_N} = \frac{12}{7 \zeta(3)} \frac{(N_1 + N_2 / z^2)^2}
{(N_1 + N_2 / z^4)(N_1 + N_2 / 2)}.
\label{44}
\end{equation}
3)For $\zeta_2 < \eta < \zeta_{c1}$:
\begin{equation}
\frac{C_S - C_N}{C_N} = \frac{12}{7 \zeta(3)} \frac{(N_1 + N_2 / z^2)^2}
{(N_1 + N_2 / z^4)(N_1 + N_2)}.
\label{45}
\end{equation}
4)For $\eta = \zeta_{c1}$:
\begin{equation}
\frac{C_S - C_N}{C_N} = \frac{12}{7 \zeta(3)} \frac{(N_1 + N_2 / z^2)^2}
{(N_1/2 + N_2 / z^4)(N_1/2 + N_2)}.
\label{46}
\end{equation}
5)For $\zeta_{c1} < \eta < \zeta_{c2}$:
\begin{equation}
\frac{C_S - C_N}{C_N} = \frac{12}{7 \zeta(3)}.
\label{47}
\end{equation}

The formulae (43)-(46) implicity demonstrate the consequence of inclusion and
turning-off of the overlapping bands as the parameter $\eta$ increases. The
expression (45) corresponds to the existence of both bands and to the maximum
value of the superconducting transition temperature.This expression coinsides
in form with the one obtained in the model with phonon superconductivity
\cite{Moskalenko_2},\cite{Moskalenko_9}.

By using the above formulae the relative jump of the electron specific heat one
has $(C_S - C_N) / C_N < 1.43$ at any values of the parameter $\eta$(or $N_0$)
and the former can achieve sufficiently small values. Note that the small
values of the quantity $(C_S - C_N)/C_N = 0.43 - 1.14$ have been considered in
the thallium ceramics \cite{Junod}.

The specific heat jump dependence on $N_0$ corresponding to $\lambda_{n m} > 0$
and to $\lambda_{n m} < 0$ is presented in fig. 8(a) and (b),respectively. In
both cases the ratio $(C_S - C_N)/ C_N$ is equal to 1.43 in the range where
only one energy band exists $(\eta < \zeta_2$ and $\eta > \zeta_{c1})$.  This
value corresponds to the BCS theory. This ratio decreases after inclusion of
both bands.In the case $\lambda_{nm} > 0$ we obtain a smooth dependence with a
slightly spread minimum. Nevertheless, in the case $\lambda_{nm} < 0$ we
observe a quick decrease of this ratio as $N_0$ increases followed by a slower
rate of growth in the range with overlapping energy bands.

The mentioned above formulas for the jump of heat capacity (42),
and (43)-(47) as well, correspond to the isotropic two-band superconductor
at $\Delta_{12}=\Delta_{21}=0$. In particular, for $\Theta_n=F_n=1$ formula
(42) gives the case of electon-phonon mechanism of superconductivity
\cite{Moskalenko_2}, \cite{Moskalenko_9} in two-band isotropic
system.  In case of anisotropic electron-phonon interaction the result is
presented in Ref. \cite{Mishonov_1}-\cite{Mishonov_3} and have succeeded to
describe the behavior of the heat capacity at  $T=T_c$ in $MgB_2$ compound.

Note that our investigations were made in the meanfield approximation, and it
is in this approximation that the proposed superconductivity theory, with two
overlapping energy bands, describes the properties of the system for an
arbitrary ratio of $T_c$ to $N_0$. The mean-field approximation
itself, however, may turn out to be insufficient when very low carrier densities
are considered.  It becomes necessary here to take into account the
fluctuations of the order parameters near the superconducting transition
temperature.Our numerical calations were made mainly for values $T_c
/\varepsilon_F \sim 10^{- 1} - 10^{-2}$(cf.the data of Figs. 7 and 9), and in
this respect we can be assumed to have a physical picture that is qualitatively
close to the real one.  To be sure, we are still faced here with the question
of the nature of the superconductivity, namely, will it be based on the Cooper-
pair production mechanism or will it be determined by the Bose condensate.Good
results in the ground state are obtained in the BCS theory for $\Delta \ll
|\mu|$, but a condensate of noninteractiong bosons is produced in the opposite
limit.Analysis of the intermediate region at finite temperature is still an
unsolved problem, and our results are an interpolation of the BCS mechanism to
this intermediate region.
\\\\
{\bf 2.5 Electron-phonon mechanism of superconductivity}

In the precious sections we has derived the equations and the analytical
expressions in particular cases for the quantities $T_{C}$ , $\mu \left(
T\right) $, and for the absolute and relative jump of heat capacity at the
point $T=T_{C}$. The equations are valid both for electron and phonon
mechanism of superconducting pairing. Formally speaking, this feature
appears in a way of cut-off while having performed the integration over
energy in the main equations (see section 2.2). The graph plots of the
presented here dependences of various quantities (fig. 1-9) correspond the
non-phonon superconductivity.

We present also the results for the electron-phonon superconductivity,
having considered as earlier the weak and strong hybridization, and chosen
the same values of parameters $\lambda _{nm}$ as in section 2.2.

We choose also the following values .$\varsigma _{1}=0$ eV, $\varsigma
_{2}=0.01$ eV, $\varsigma _{c1}=0.02$ eV, $\varsigma _{c2}=0.03$ eV. The
results of numerical calculations are presented on the fig.1 - 10. The shape
of dependences of the quantity $T_{C}$ on $N_{0}$ presented on the fig.1 and
10 is easily observed not to be the same. This difference appears clearer
especially in the case of weak hybridization. In the case of strong
hybridization there is a bell-shape dependence of the quantity $T_{C}$ on
the density of charge carriers.

The kink of chemical potential $\mu \left( T\right) $ at the point $T=T_{C}$
(fig. 11) is well distinguished and disappears at the values $\mu \sim 6$
meV what is as in the case of electron pairing mechanism (fig. 2) achievable
to the experimental observation.

The dependence of the jump of electron heat capacity $\left(
C_{S}-C_{N}\right) |_{T=T_{C}}$ on the density of charge carriers differs
distinctly at the weak hybridization as well (compare fig.6a and 12). In the
case of phonon pairing mechanism of superconductivity and weak hybridization
the maximum of the quantity $\left( C_{S}-C_{N}\right) /\,VN_{1}$ is shifted
towards low densities of charge carriers in comparison with the non-phonon
pairing mechanism. The curves (fig.6a and 12) that correspond to the strong
hybridization differ as well. The dependence of the relative jump of
electron heat capacity on the density of charge carriers in both cases of
hybridization has complex shape ( see fig.6b and 13) and depends essentially
upon theory parameters. Regarding the value of $N_{0}$ the quantity $\left(
C_{S}-C_{N}\right) /\,C_{N}$ at $T=T_{C}$\ can be as greater as lower than
1.43 (the value that corresponds to the isotropy BCS model with phonon
pairing mechanism of superconductivity). Detailed analysis of these results
and their comparison with experimental data might shed light upon the
mechanism of superconductivity in two-band system.

As was noticed above the discovery of high-$T_{C}$ superconductivity in
magnesium boride $MgB_{2}$ $T_{C}\sim 39K$ \ has lead to rapid development
in experimental and theoretical studies. Confirmation of two-band nature of
this system (presence of two energy gaps) \cite{Ponomarev} and electron-phonon
mediated pairing mechanism of superconductivity \cite{Bud'ko} - \cite{Choi_3} is
very important discovery. The singularities of two-band system in temperature
dependence of heat capacity, penetration depth of magnetic field, upper
critical magnetic field and other physical quantities are observed. The
background of all these studies is the model of Moscalenco that was
suggested by him in 1959 \cite{Moskalenko_2}. On the basis of this model
Moscalenco and his colleagues have made a great number of studies long before
the discovery the compound $ MgB_{2}$. Many modern authors to our pity
represent their own results as originally new ones not making any references on
the pioneer works of  Moscalenco and his colleagues (see  \cite{Palistrant_1},
\cite{Moskalenko_8}).

In order to describe the physical properties of $MgB_{2}$ it is necessary to
determine the parameters of theory on the basis of present experimental
data, to take into account anisotropy of matrix element of electron-phonon
interaction \cite{Mishonov_1} - \cite{Mishonov_3}, important role of Van
Hove singularities \cite{Loram}, it is apt to take into consideration strong
electron-phonon interaction and other singularities as well as multi-band
nature of the system.

Because parameters of theory in the two-band model presented in this article
are unknown, our studies performed here permit us to make qualitative
conclusions about superconducting properties of $MgB_{2}$.

1. The quantity $T_{C}$ as a function of the density of charge carriers has
bell-shape dependence. At small hybridization there is wide field of
densities of charge carriers $N_{0}$ at which $T_{C}$ depends slightly upon $
N_{0}$. At great hybridization the maximum in the dependence of $T_{C}$ on $
N_{0}$ is notably distinguished. The height and position of this maximum is
determined by relationship $N_{1}/N_{2}$ as well as by values of coupling
constants $\lambda _{nm}$.

2. The chemical potential at the point $T=T_{C}$  has kink at low densities
of charge carriers. This kink disappears with increasing $N_{0}$. The
overlapping of energy bands favors the experimental confirmation of this
kink.

3. The positions of maximums of $T_{C}$ and difference $\left(
C_{S}-C_{N}\right)$ at $T=T_{C}$ as a functions of density of charge carriers
do not coincide.

4. The relative jump of electron heat capacity $\left( C_{S}-C_{N}\right)
/C_{N}$ at the point $T=T_{C}$ depends on the density of charge carriers and
can get as great (greater than 1.43) as small (less than 1.43) values. The
more detailed study of the two-band model considered in this article with
reduced density of charge carriers and account of all possible pairings of
electrons at phonon mechanism of superconductivity can be found in the
second article of the Ref. \cite{Kochorbe_1}.

Note that in the most of works about $MgB_{2}$ theoretical and experimental
studies of superconducting properties are made at given value of density of
charge carriers. But there are studies (see for example,\cite{Akis_1},
\cite{Akis_2}) where the dependence of the quantity $T_{C}$ on the density
of charge carriers is investigated on the basis of simplified Moscalenco
model assuming $ V_{11}=V_{22}=0$, $V_{12}\neq 0$ . In order to achieve high
values of $T_{C}$ it was assumed $V_{12}>1$.

The change of density of charge  carriers can be done by doping as it was
suggested in works \cite{Mishonov}, \cite{Rodriguez} where compounds
$MgB_{2-x}C_{x},Mg_{1-y}D_{y} B_{2}(D=Li,Al\;{\rm and\;others})$.  have been
studied. Having introduced impurity in such compounds, the interband scattering
of electrons on a impurity, that leads to the decrease of the quantity $T_{C}$
with increase of impurity concentration \cite{Lagos}, occurs as well as the change
of chemical potential. So, in order to confirm the properties 1 - 4 in the
experimental perspective we have to exclude the interband impurity scattering,
to neutralize it somehow. It can be done , for example, by introducing the
impurity outside the plane that is responsible for superconductivity.
Therefore, in this plane the arrangement of electrons is preserved and the
scattering on the impurity potential is absent \cite{Lagos}.\\

{\bf 2.6 Conclusion}

We have developed a superconductivity theory for a system with two overlapping
energy bands on the Fermi surface.This theory is valid in the weak-field
approximation for any carrier density, including a low one $(\mu \sim T_c)$.

The main results are the following.

1. We have introduced a system Hamiltonian (1) which can account for
superconducting pairing of electrons both within each band and from different
bands. A system of self-consistent equations was derived for the order
parameter $\Delta_{nm}$ (2) and for the chemical potential $\mu$ (6).

2. We have used a sub-band representation, in which the basic equations at
temperatures close to critical can be expressed as the set (16)-(19) for the
four-band model. This set can be used to consider both the phonon and
non-phonon superconductivity mechanisms and can be used to describe
superconductivity in a system with low carrier density $(\mu \sim T_c)$.The low
carrier density notwithstanding, this system can lead  to rather high
$T_c$ in view of the inclusion of more interband interactions than in
Ref.\cite{Moskalenko_2}, which are connected with formation of superconducting
pairs of electrons from different bands.

3. Our model offers more possibilities of describing various two - band
systems, since a major role is played in the theory by the ratio $N_2 / N_1$
of state densities of electrons from different bands, as well as by interation
constants $\lambda_{nm}(n; m = 1 - 4)$. The foregoing is in fact clearly
demonstrated in Figs.1 and 6a which show respectively the dependences of the
critical temperature $T_c$ and of the electronic heat capacity $(C_S - C_N)_{T
= T_c}$ on the carrier density for a nonphonon superconductivity mechanism and
strong (curves 4 - 5) hybridization.

4. The plot of the chemical potential of a superconductor with low carrier
density versus temperature has at the point $T = T_c$ a sharp kink that becomes
less peaked and vanishes when the carrier density is increased. It vanishes at
$\mu \approx 2$ meV in the BCS model \cite{Van} and at $\mu \approx 8$ meV in
the two - band case (Fig. 2). Overlap of the energy band produces thus
favorable conditions for revealing anomalies in the $\mu (T)$ dependence and
hence for elucidating the superconductivity mechanisms.

5. Analytic expressions were obtained for the absolute $(C_S - C_N)$ (37)
and relative $(C_S - C_N) / C_N$ (40) jumps of the  electron heart capacity at
$T = T_c$. The behavior of $(C_S - C_N)$ as a function of the carrier density
$N_0$ does not, generally speaking, duplicate the behavior of $T_c (N_0)$ (see
Figs. 1 and 6a) and is determined by the parameters of the  theory. Situations
are possible in which $T_c$ and $(C_S - C_N)_{T = T_c}$ have maxima at one and
the same density $N_0$ (e.g., curves 3 - 5). The character of the dependence of
the ratio $(C_S -  C_N)/C_N$ at the point $T = T_c$ on $N_0$ is also determined
by the parameters of the theory and depend substantially on the carrier
density. This quantity can be either large or small:  $(C_S - C_N) / C_N >
1.43$ or $(C_S - C_N) / C_N < 1.43$.

6. At large carrier densities $(\mu \gg T_c)$ we have $\Delta_{12}, \Delta_{21}
\ll \Delta_{11}, \Delta_{22}$, so that simpler equations for $T_c, \Delta_{nm},
(C_S - C_N)_{T = T_c}$ and $(C_S - C_N)/ C_N$ can be obtained by putting
$\Delta_{12} =\Delta_{21} = 0$ in (2) and $N_{3} = N_{4} = 0$ in (19),
(37) and (40).  The ensuing results agree with the corresponding equations for
phonon \cite{Moskalenko_2}, \cite{Moskalenko_9} and non - phonon
\cite{Palistrant_2} superconductivity mechanisms.

We have considered two cases: (a) all constants both for the interband and for
the intraband interaction are positive: ($\lambda_{nm} > 0$, this corresponds
to effective attraction between carriers); (b)$\lambda_{nm} < 0$, (this
corresponds to repulsion). We suppose that case (2) corresponds to the strong
Coulomb interaction when any attraction between carriers is suppressed. The
interband interaction coupling constants which are involved in the $T_c$ square
definition are here the unique reason for generating a superconductivity
mechanism, with the intraband interactions preventing one. Nevertheless, high
values of $T_c$ can be achieved provided
$\lambda_{11}\lambda_{22}<\lambda_{12}\lambda_{21}$.  In the case of
$\lambda_{nm}>0$ the high values of $T_c$ are easily acquired without imposing
any restrictions on the values of the parameters $\lambda_{nm}$. The results
concerning the dependences of $2\vert\Delta_n(0)\vert/T_c$ on the carrier
concentration are of great interest to us. These dependences can be
interpreted as the dependences on the oxygen content in the yttrium ceramics
or on the doping content in the lanthanum ones.  The supplemental dependence on
the carrier concentration results in our model in a difference from that with a
two-band phonon superconductivity mechanism \cite{Moskalenko_2},
\cite{Moskalenko_9}.

The theory, represented in this work, gives the essential dependence of the
relative specific heat jump $(C_s - C_N)/C_N$ on the carrier concentration.
In particular, sufficiently small values of this ratio are possible (see
fig. 6a and 6b).
As has been noted before, small values of this quantity have been observed,
for example, in the thallium ceramics \cite{Junod}. So the specific heat
jump dependence on the carrier concentration which has a maximum (see fig. 7)
is of great interest to us. The value  of $N_0$ which corresponds to the above
maximum coincides with the one which defines the maximum of the
superconductivity transition temperature. The specific heat dependence on the
doping concentration has been observed, for instance, in $La_{2-x}Sr_xCu_2O_4$
\cite{Hongshum}.

In the case of electron-phonon mechanism of superconductivity the results
obtained in this work can be used to describe the thermodynamic properties
of superconducting $MgB_{2}$ compound in all scope of densities of charge
carriers (see section 2.5). In this way we have to determine the parameters
of theory for this compound and find suitable experimental technique to
lower the density of charge carriers preferably up to values where $\mu \sim
T_{C}$ . As it was shown in many works the superconductivity in $MgB_{2}$
compound with its inherent density of charge carriers is described by
Moscalenco two-band model \cite{Moskalenko_2}.

\begin{center}
\bf{3. SUPERCONDUCTIVITY IN TWO-BAND SYSTEM PASSED TO THE STATE OF A BASE
CONDENSATE OF LOCALIZED PAIRS AT T = 0 }.
\end{center}

{\bf 3.1. Analitical solutions for the basic equations in the meanfield
approximation at T = 0}.

The two-band superconducting system with  an arbitrary carrier concentration
is described by the Hamiltonian (1). Using this Hamiltonian and the Green
functions method \cite{Abricosov_1}, we have obtain the self consistent
system of equations for the order paramiters $\Delta_{nm}(n; m = 1,2)$ and the
chemical potential $\mu\,(2),(6)$. These equations we will study at $T = 0$.

Since this system of equations cannot be solved in the general form, our goal is
to find these solutions for the region of low carrier concentrations when the
system passes to the state of Bose condensate of localized pairs
\cite{Leggett_2}, \cite{Randeria} $(\mu_n < 0)$ at $(\Delta_{mn}/
\mu_{n,m})^2 \ll 1$.  For these conditions, in expressions (2)-(6),
expansion in terms of $\Delta_{nm}^2$ can be performed.

For the sake of simplicity, redefine the indices $11 \rightarrow 1; 22
\rightarrow 2; 12 \rightarrow 3$. As a result, the system of equations (2) and
(6) can be rewritten as a system of equations for the order parameters of the
three-band model \cite{Palistrant_6}.Limiting ourselves to terms of order
$\Delta_{n}^{2}$, we obtain the system of equations for finding the values ot
the chemical potential $\mu$ and order parameters $\Delta_n (n = 1, 2, 3)$.

This system of equations and theirs solutions have study in paper
\cite{Palistrant_7}. As follows from this paper the solutions for $\mu$ and
$\Delta_{n}^{2}$ are rather cumbersome and contain a complicated  dependence on
the carrier concentration and the intra- and interband interaction constants.
To clarify the role or the overlapping energy bands, some particular cases will
be considered below. Let us consideder two simpler cases here.

A. The Moskalenko model \cite{Moskalenko_2} assumes the existence of
intraband $V_{11}$ and $V_{22}$ and interband $V_{12}$ electron-electron
interactions corresponding to electron pairing inside one band and transitions
of pairs as a whole from one band to the other. In thise case, $\Delta_3 = 0$
and in the region of a small concentrations $(\mu < 0, \Delta_{n}^{2}/ \mu^2
\ll 1)$ we reduce the equations for $\Delta_1, \Delta_2$ and $\mu$ to the
following form:
\begin{equation}
\mu (1 - \exp \alpha_1) + \frac{\Delta_{1}^{2}}{4} f_1 =\tilde
\zeta_{c_1},\,\,\,\,\,\mu (1 - \exp \alpha_2) + \frac{\Delta_{2}^{2}}{4} f_2
=\tilde \zeta_{c_2},\,\,\,\,\,n_1 F_1 \frac{\Delta_{1}^{2}}{4} + n_2 F_2
\frac{\Delta_{2}^{2}}{4} = \tilde n,
\label{48}
\end{equation}
where
$$
\alpha_1 = \frac{V_{22} - V_{12} \Delta_2 / \Delta_1}{N_{1}^{'}(V_{11}V_{22} -
V_{12}V_{21})},\,\,\,\,\,\,\,\,\, \alpha_2 = \frac{V_{11} - V_{21} \Delta_1 /
\Delta_2}{N_{1}^{'}(V_{11}V_{22} - V_{12}V_{21})},
$$
\begin{equation}
N_{n}^{'} = \frac{N_n}{2},\,\,\,\,\tilde n = \frac{N}{2(N_1 + N_2)},\,\,\,\,
n_n = \frac{N_n}{N_1 + N_2},
\label{49}
\end{equation}
$$
f_n = \frac{(1 - \exp \alpha_n)(1 - \exp 2 \alpha_n)}{\zeta_{cn} \exp
\alpha_n},\,\,\,\,F_n = \frac{(1 - \exp \alpha_n)^2}{\zeta_{cn} \exp \alpha_n},
$$
$$
\tilde \zeta_{cn} = \zeta_{cn} + \zeta_n(1 - exp \alpha_n),
$$
and where $N_n$ is the density of electron states in the $n$th sheet of the
Fermi surface and $\zeta_{cn}$ is the cutoff energy for the integrals. The
solutions of. Eqs.(48) have the form
\begin{equation}
\mu =\left(n_1 \tanh \frac{\alpha_1}{2} + n_2 \tanh
\frac{\alpha_2}{2}\right)^{-1}\,\Biggl\{-\frac{1}{2}\left[n_1 \tilde \zeta_{c_1}
\left(1 - \tanh \frac{\alpha_1}{2}\right) + n_2 \tilde \zeta_{c_2}\left(1 -
\tanh \frac{\alpha_2}{2}\right)\right] + \tilde n\Biggl\},
\label{50}
\end{equation}
$$
\Delta_{1}^{2} = \frac{\zeta_{c_1}}{sinh \frac{\alpha_1}{2}\,\cosh\,
\frac{\alpha_1}{2}} \left( n_1\,\tanh\, \frac{\alpha_1}{2} + n_2\,\tanh\,
\frac{\alpha_2}{2}\right)^{-1} \times
$$
\begin{equation}
\times \Biggl\{\tilde n + n_2 \left[\zeta_{c_1}\tanh \frac{\alpha_2}{2}\left
(\coth \frac{\alpha_1}{2} - 1 \right) + \zeta_{c_2} \left(\tanh
\frac{\alpha_2}{2} - 1 \right)\right]\Biggl\}.
\label{51}
\end{equation}
The expression for $\Delta_2$ can be obtained from Eq.(51) by replacing indices
$1 \leftrightarrow 2$.On the basis of Eq.(51) we obtain the equation for the
ratio $\Delta_1 / \Delta_2$,
\begin{equation}
\left(\frac{\Delta_1}{\Delta_2}\right)^2 =\frac{\zeta_{c_1}\tilde n + n_2
\zeta_{c_1} \left[\zeta_{c_1} \tanh \frac{\alpha_2}{2}(\coth
\frac{\alpha_1}{2} - 1) + \zeta_{c_2}(\tanh \frac{\alpha_2}{2} - 1)\right]
\sinh \alpha_2}{\zeta_{c_2}\tilde n + n_1
\zeta_{c_2} \left[\zeta_{c_2} \tanh \frac{\alpha_1}{2}(\coth
\frac{\alpha_2}{2} - 1) + \zeta_{c_1}(\tanh \frac{\alpha_1}{2} - 1)\right]
\sinh \alpha_1}.
\label{52}
\end{equation}

Let us pass to the limit of symmetrical bands. In this case,
$\alpha_1=\alpha_2$, $\zeta_{c_1}=\zeta_{c_2}$, $n_1 + n_2=1$ and formulas
(51) transform into
\begin{equation}
\mu=-\frac{\tilde \zeta_{c_1}}{2}\left[\coth\frac{\alpha_1}{2} - 1\right] +
\tilde n\coth\frac{\alpha_1}{2},\ \ \ \Delta_1^2=\Delta_2^2=\frac{\tilde n
\zeta_{c_1}}{\sinh^2\alpha_1/2}.
\label{53}
\end{equation}
Formulas (53) coincide in form with analogous expressions for one-band
superconductors \cite{Gordar}, and the difference is in the determination
of $\alpha_1$ (49), including interband interactions in addition to intraband
interactions.  This fact leads to a decrease in the denominator in the second
formula in (53), and consequently, to an increase in $\Delta_1$, in comparison
with the one-band case. At the same time, a more essential shift of $\mu$
towards the negative region occurs under the condition of weak changes in the
value of $\zeta_{c_1}$. As follows from (51), in the case of an anisotropic
two-band system, the formulas are modified due to the difference in the
parameters of the considered systems. So, for example, in the determination of
$\Delta_1$ for the carrier concentration $\tilde n$, besides renarmalization of
the coefficient, an additional term arises due to the presence of the second
band ($n_2 \not = 0$).

As follows from formulas (50), (51), the dependence of $\mu$ and $\Delta_n$
on the carrier concentration has a complex character due to the ratio
$\Delta_1/\Delta_2$ entering the expressions determining $\alpha_1$ and
$\alpha_2$. The $\Delta_1/\Delta_2$ ratio itself is determined by Eq. (52).
From this formula it is easy to see that for a weakly anisotropic system (the
bands are nearly similar), the dependence of $\Delta_1/\Delta_2$ on the carrier
concentration  can be neglected. As a result, $\mu$ and $\Delta_n$ are
determined by the explicit dependence on $\tilde n$ (51), as in the case of a
one-energy band. In the case of a strongly anisotropic system, the dependences
of $\mu$ and $\Delta_n$ must be determined from formulas (50), (51) and
equation (52). The result can be esentially different from the case of one-band
superconuctors.

In the behavior of a superconducting system with a small carrier concentration,
the existence of a two-particle bound state plays an important role
\cite{Leggett_2}, \cite{Gorbatsevich_1}.  The equation for determining
the binding energy $\varepsilon_b$ in a two-band system is obtained by
considering the two-particle "density-density" Green's function at
$\varepsilon_F=0$ and the corresponding equation for the vertices
\cite{Palistrant_6}, \cite{Palistrant_8},
$$
\Gamma_{nm} = - V_{nm} + \sum_{n_1} V_{nn_1} \xi_{n_1}(|\varepsilon_b|)
\Gamma_{n_1 m}.
$$
\begin{equation}
\xi_n (|\varepsilon_b|) = \sum_{\vec k}\left[2 \varepsilon_n (\vec k) +
|\varepsilon_b| \right]^{-1}.
\label{54}
\end{equation}

This system is written for the case of $\Delta_n = 0$ and $\varepsilon_F = 0$
and corresponds to the two - particle interaction without accounting for
multiparticle effects.The bound state of two particles arises when the vertex
diveges. Consequently, on the basis of Eq.(54), the following equation for the
bound-state energy $\varepsilon_b$ is obtained:
\begin{equation}
1 - V_{11} \xi_1 (|\varepsilon_b|) - V_{22} \xi_2 (|\varepsilon_b|) +
(V_{11}  V_(22) - V_{12 }^2) \xi_1(|\varepsilon_b|) \xi_2 (|\varepsilon_b|)
= 0.
\label{55}
\end{equation}
From the condition $N_0 \rightarrow 0$, we obtain \cite{Palistrant_9}
\begin{equation}
1 - V_{11}  \xi_1 (2 \eta) - V_{22} \xi_2 (2 \eta) +
(V_{11} V_(22) - V_{12 }^2) \xi_1(2 \eta) \xi_2 (2 \eta) = 0.
\label{56}
\end{equation}
Comparing the two last equations, we can state that as $N_0 \rightarrow 0$,
\begin{equation}
|\varepsilon_b| = 2 \eta,\,\,\,\,\,(\eta = - \mu).
\label{57}
\end{equation}

B. All the interaction constants are equal to zero $(V_{nm}^{ke} = 0)$,
except $V_{12}^{V_{21}}$. In this case $\Delta_1 = \Delta_2 = 0$ and we
have the following expression for $\Delta_3$ and $\mu$:
\begin{equation}
\Delta_{3}^{2} = \Delta_{12}^{2} = \frac {\tilde n (1 + m_1 / m_2)
(\zeta_{c_1} + \zeta_{c_2})}{8 n_1 \sinh^2 \beta_1 / 2}
\label{58}
\end{equation}
\begin{equation}
\mu = - (\zeta_{c_1} + \zeta_{c_2}) \frac{1}{4} \left(\coth
\frac{\beta_1}{2} - 1 \right) + \frac{\tilde n (1 + m_1 / m_2)}{4 n_1}\coth
\frac{\beta_1}{2},
\label{59}
\end{equation}
Where $m_n$ is the effective mass of the electron in $n$ - band.
\begin{equation}
\zeta_{c_2} = \frac{m_1}{m_2} \zeta_{c_1},\,\,\,\,\,\beta_1 = \frac{1 + m_1
/ m_2}{4 N_{1}^{'} V_{12}^{21}}.
\label{60}
\end{equation}
If we assume that the bands are symetric, i.e., $\zeta_{c_1} =
\zeta_{c_2}$,$m_1 = m_2, n_1 = 1 / 2$, the above formulas are modifided
into the expressions
\begin{equation}
\mu = - \zeta_{c_1}\frac{1}{2} \left(\coth \frac{\beta_1}{2} - 1 \right)
+ \tilde n \coth \beta_1 / 2 ,\,\,\,\,\,\Delta_{3}^{2}
=\frac{\zeta_{c_1} \tilde n}{\sinh^2 \beta_1 / 2},
\label{61}
\end{equation}
whose form resembles (53).

The above considered case of the presence of one order parameter
($\Delta_{12}$) in the system leads to a concentration dependence analogous to
the case of symmetrical bands.

For a system with Hamiltonian (1), the expression for the difference in
thermodynamic potentials for the states with $\Delta_{nm} \not = 0$ and
$\Delta_{nn} = 0$ can be written as (35). Using the calculation technique
reported in \cite{Palistrant_2} and the system of equations for
$\Delta_n$, we can reduce the difference (35) to the form
\begin{equation}
\frac{\Psi (\Delta) - \Psi (0)}{V} = \frac{1}{8} N_{1}^{'}\left[ f_{1}
\Delta_{1}^{4} + f_2 \,\Delta_{2}^{4} \frac{N_2}{N_1} + 2 f_3\,\Delta_{3}^{4}\,
\right],
\label{62}
\end{equation}
where $f_n < 0$ \cite{Palistrant_7}. As a concequence, the state with $\Delta_n
\not = 0$ is more advantageous  than the trival solution $\Delta_n = 0$. The
advantage increases as the number of order parameters increases in connection
with accounting for all kinds of inter- and intraband interactions.

In the case of symetrical bands, expression (62) can be rewritten in the form
\begin{equation}
\frac{\Psi (\Delta) - \Psi (0)}{V} = -2 N^{'}\coth \frac{\alpha_1}{2}\,\tilde
n^2
\label{63}
\end{equation}
This formula coincides in form with an analogous expression for the one - band
superconductor \cite{Gordar}.

{\bf 3.2 Application of the path integral method.}

In this section the path  integral method as applied to the two - band model is
developed  and, on this basis, the procedure for transition from the Fermi to
Bose elementary exitations is given.

We start from following Hamiltonian describing the two-band systems:
$$
H = \int\,dr\,\sum_{n \sigma} \Psi_{n \sigma}^{+}(\tau) H_{0n} \Psi_{n \sigma}
(\tau) - \frac{1}{2} \sum_{nm} V_{n m} \int dr \sum_{\sigma\,\sigma^{'}}
$$
\begin{equation}
\Psi_{n \sigma}^{+}(\tau)\Psi_{n \sigma^{,}}^{+} (\tau) \Psi_{n
\sigma^{'}} (\tau) \Psi_{n \sigma}(\tau)(1 - \delta_{\sigma \sigma^{'}}),
\label{64}
\end{equation}
where the first term corresponds to the kinetic energy, the second is
responsible for the superconductivity, the band indices $n; m = 1, 2, V_{nm}$
are the constants of intra- and interband interaction, $\Psi_{n
\sigma}^{+}(\tau)$ is the particle creation operator in the band $n$ with
spin $\sigma$, and $H_{0n} = \frac{\bigtriangledown^2}{2 m_n} - \mu_n$.
Hamiltonian (1) correspond to expression of the Moskalenko model
\cite{Moskalenko_2} considering only intarband pairind and the transitions of
Cooper pairs as a whole from one band to another.

The expression for the statistical sum of a two-band superconductor after
introducting additional scalar fields $\Delta_n$, Hubbard-Stratonovich
transformations, integrations under Fermi fields, and transition into $n k
\Omega$ representation has the following form \cite{Palistrant_10} (the
detailed version see \cite{Palistrant_9}):
\begin{equation}
Z = C \int \left(\prod_{i = 1,2} d \Delta_{i}^{*} d \Delta_{i}\right) \exp {-
S_{eff}^{(2)}},
\label{65}
\end{equation}
where
$$
S_{eff}^{(2)} = \frac{1}{\beta}\,\sum_{nm}\,\sum_{\vec q
\omega}\,\Delta_{n}^{*} (\vec q \omega)
$$
\begin{equation}
\biggl\{ V_{nm}^{-1} +  \Pi_n(\vec q\omega)\delta_{nm}\biggl\} \Delta_m (\vec
q \omega),
\label{66}
\end{equation}
\begin{equation}
\Pi_n (\vec q, \omega) = \frac{1}{\beta}\sum_{\vec k \Omega} G_{- n} (\vec k
\Omega) G_{+ n} \left(\vec k -  \vec q, \Omega - \omega \right)
|\chi_n (\vec k, \vec k - \vec q)|^2,
\label{67}
\end{equation}
\begin{equation} G_{\pm
n}^{-1}(k \Omega) = - i \Omega \pm \frac{k^2}{2 m_n} \mp \mu_n.
\label{68}
\end{equation}

Here $V_{nm}^{- 1}$ is the matrix element of inverse to the interaction matrix

$$ \hat V = \left( \begin {array}{ccc}
	       V_{11} & V_{12}\\
	       V_{21} & V_{22}
	       \end {array} \right).     $$

and $\chi_n (\vec k, \vec k - \vec q)$ are the integrals over the elementary
cell under the Bloch functions.

Further calculations are connected as in the case of a single band
\cite{Gorbatsevich_1} with the existence of a nonzero solution of Eq. (55),
determining the energy of the  bound state $\varepsilon_b$, and we consider
the case $T  \ll |\varepsilon_b|$.

In the mean-field approximation $\mu = \varepsilon_b / 2$ (57).Let us suppose
that $\mu$ is near $\varepsilon_b / 2$. After integrating under $\Omega$ in
formula (67) we make the expansion under $\omega, \vec q$ and the difference
$(2 \mu - \varepsilon_b)$.  After that we acquire
\begin{equation}
\Pi_n (\vec q, \omega) = -\xi_n (|\varepsilon_b| ) + \gamma_n \left[ - i \omega
+ \frac{q^2}{2 m_{n}^{*}} - \mu^{*}\right] ,
\label{69}
\end{equation}
were
\begin{equation}
\gamma_n =- \frac{\partial \Pi_n}{\partial(2 \mu)}\,\,\,\,\mu^{*} = 2 \mu -
\varepsilon_b,\,\,\,\frac{1}{m_{n}^{*}} = \gamma_{n}^{- 1} \frac{\partial^{2}
\Pi_n} {\partial \vec q^2}.
\label{70}
\end{equation}

Let us substitute (66) into (65) and perform the functional integration under
additional Bose fields, leading to the canonical form by orthogonal
transformation. We obtain
\begin{equation}
\ln Z = - \sum_{\vec q \omega} \ln \biggl\{1 + V_{11} \Pi_1 (\vec q, \omega) +
V_{22}\Pi_2 (\vec q, \omega) + \left(V_{11} V_{22} - V_{12}^{2}\right) \Pi_1
(\vec q, \omega) \Pi_2 (\vec q, \omega) \biggl\}.
\label{71}
\end{equation}
We substitute in this formula expressions for the $\Pi_n (\vec q, \omega)$ [see
Eq.  (69)] and take into account Eq. (55), determining the energy of the bound
state $\varepsilon_b$, and also relationships
\begin{equation}
V_{11} - \left(V_{11} V_{22} - V_{12}^{2} \right) \xi_2 = V_{21} \Delta_1 /
\Delta_2,\,\, V_{22} - \left(V_{11} V_{22} - V_{12}^{2} \right) \xi_1 = V_{12}
\Delta_2 / \Delta_1,
\label{72}
\end{equation}
resulting from the system of equations for the order parameters $\Delta_n$.
Thus we obtain
\begin{equation}
\ln Z = -\sum_{\vec q \omega} \ln \Biggl\{ \gamma_1 V_{21}
\frac{\Delta_1}{\Delta_2} \left(- i \omega + \frac{q^2}{2 m_{1}^{*}} - \mu^*
\right) + \gamma_2 V_{12} \frac{\Delta_2}{\Delta_1} \left( - i \omega +
\frac{q^2}{2 m_{2}^{*}} - \mu^* \right)\Biggl\}.
\label{73}
\end{equation}
For the charge carrier density we have
\begin{equation}
n = \frac{1}{\beta}\,\frac{\partial}{\partial \mu^*} \ln Z = \frac{1}{\beta}
\sum_{\vec q \omega} \left(- i \omega + \frac{q^2}{2M^*} - \mu^{*} \right)^{-1}
= \sum_{\vec q} \left(\exp \left[ \beta \left( \frac{q^2}{2 M^*} - \mu^*
\right)\right] - 1 \right)^{-1},
\label{74}
\end{equation}
were
\begin{equation}
(M^*)^{-1} = \frac{1}{m_{1}^{*}} \left(\gamma_1 \frac{\Delta_1}{\Delta_2} +
\gamma_2 \frac{\Delta_2}{\Delta_1} \frac{m_{1}^{*}}{m_{2}^{*}}\right)
\left(\gamma_1 \frac{\Delta_1}{\Delta_2}+\gamma_2 \frac{\Delta_2}{\Delta_1}
\right)^{-1}.
\label{75}
\end{equation}
Expression (75) has been obtained in the  approximation in which only the
quadratic part of complete action is taken into account. In this approximation
we obtain the ideal Bose gas with effective mass $M^*$ and chemical potential
$\mu^* = 2 \mu - \varepsilon_b$.In distinction from the case of a
single band \cite{Gorbatsevich_1}, here $\varepsilon_b$ and $\mu$ are
determined on the two-band basis (55), (50) and $M^*$ is defined by (75). In
the case where the effective masses from different bands are equivalent
$(m_{1}^{*} = m_{2}^{*})$, the relationship $M^* = m_{1}^{*} = m_{2}^{*} $
results from the last formula.  If the energy band with effective mass
$m_{1}^{*}$ overlaps with the wider band $(m_{1}^{*} >m_{2}^{*})$, in the field
with low charge carrier density a lighter Bose gas arises $(M^* < m_{1}^{*})$
in comparison with the case of a single-band superconductor effective electron
mass of which is equal to $m_{1}^{*}$.

The system undergoes transition to the state of Bose condensation at the point
$\mu^* = 0$ and this relationship is the condition for determining the
condensation temperature $T_k$.After summation under $\vec q$ in (74) at $\mu^*
= 0$, we obtain
\begin{equation}
T_k = a_D \frac{n^{2 / D}}{M^*},
\label{76}
\end{equation}
where $D$ is the dimension of system and $a_D$ is a constant. We see that
$T_k$ depends on the charge carrier density according to the law $n^{2 / D}$ as
in the case of a single - band superconductor but only for systems with
equivalent bands $(M^* = m_{1}^{*} = m_{2}^{*})$. If the system is strongly
anisotropic $(m_{1}^{*} \not = m_{2}^{*})$, the quantity $M^*$ has the
additional inexplicit dependence  on the charge carrier density through the
same dependence of the relation $\Delta_1 / \Delta_2$. From the formulas (75)
and (76) we see that the existence of a second wider energy band can lead to
an increase in the condensation temperature $T_k$ in comparison with the case
of a single energy band. The expression for the effective action (66) can be
considered as an improved mean-field approximation, which can be also applied
at $T \not = 0$.

Taking into account the residual interaction between bosons, the effective
action has the following form:
\begin{equation}
S_{eff} = S_{eff}^{(2)} + S_{eff}^{(4)}+...       ,
\label{77}
\end{equation}
where
\begin{equation}
S_{eff}^{(4)} = \frac{1}{\beta}\,\sum_{\vec q \omega}\biggl\{|\Delta_1|^4 \pi_1
+|\Delta_2|^4 \pi_2\biggl\}   ,
\label{78}
\end{equation}
\begin{equation}
\pi_n = \frac{1}{2 \beta} \sum_{\vec k \Omega}\left[ G_{+ n}(k \Omega)\right]^2
\left[G_{- n}(k \Omega)\right]^2 = \sum_{\vec k} \left[2 \varepsilon_n (\vec k)
+ \varepsilon_b \right]^{- 3}.
\label{79}
\end{equation}
After renormalization of the Bose fields
\begin{equation}
\Delta_n = \left(\frac{\partial \Pi_n}{\partial (2|\mu|)}\right)^{- 1/2}
\varphi_n
\label{80}
\end{equation}
we have
\begin{equation}
S_{eff}^{(4)}= \frac{1}{\beta}\sum_{n}\,\sum_{\vec q
\omega}|\varphi_{n}(\vec q \omega)|^2 \pi_{n}^{'}(\vec q,
\omega),
\label{81}
\end{equation}
\begin{equation} \pi_{n}^{'} =  \left(\frac{\partial
\Pi_n}{\partial(2|\mu|)}\right)^{- 2}\pi_n.
\label{82}
\end{equation}
For the quantity $\pi_{n}^{'}$ determining the potential of two-particle
interaction in the  $n th$ band we obtain
\begin{equation}
\pi_{n}^{'}\sim m_n^{-D / 2}|\varepsilon_b|^{1 - D / 2}
\label{83}
\end{equation}

This quantity depends essentially on the system dimension and is negligible
in the case of three-edimensional. This conclusion consides with the case
of a single band  \cite{Gorbatsevich_1}. Thus , omiting the residual
interaction, which is equivalent to the mean-field approximation, is valid only
for three-dimensional systems.
\\\\
{\bf 3.3 Conclusions and discussion}

In the section 3.1 the theory of superconductivity for a two - band system
with a low carrier density at $T = 0$ was constructed. All types of
electron-electron interband and intraband interactions were accounted for . As
a result, three diferent order parameters, $\Delta_{11}, \Delta_{22}$, and
$\Delta_{12}$,\cite{Palistrant_7} arise in the sistem. The sistem of
equations for the order parameters and the chemical potential $\mu$ was
obtained. The state of the system in the region of low carrier concentrations,
where $\mu_n < 0 \,\,(n = 1,2)$ (corresponding to a Bose condensate of
localized pairs), was considered. Under the assumption that
$\Delta_{nm}^{2}/\mu_{n,m}^{2} \ll 1 $, the expansion in terms of $\Delta_
{nm}^{2}$ was conducted. This allowed the analytical expressions for $\mu$ and
$\Delta_n$ to be obtained. The expression for the difference in thermodynamic
potentials for $\Delta_{nm} \not = 0$ and $\Delta_{nn} = 0$ was also obtained.

On the basis of these investigations the following conclusions are made:

1. The overlaping of energy bands on  the Fermi surface in a system with low
carrier density gives rise  to three order parameters. In the limit of the
symmetrical bands $(\Delta_{11} = \Delta_{22} = \Delta_{12})$, the values of
these parameters are higher than in the case of the one-energy band, i.e., the
overlapping of energy bands is favorable to setting  up the superconducting
properties. At the same time, the value of $\mu$ shifts towards the negative
value region, accelerating the transition of the system from the BCS pattern to
the mode of a Bose condensate of localized pairs.

2. The order parameters $\Delta_{nm}$ and the chemical potential $\mu$ are
complex functions of inter- and intraband interaction constants, as well as of
carrier concentrations. In the weak anisotropy case (the bands are nearly
symmetrical) the dependences of $\Delta_{nm}$ and $\mu$ on the carrier
concentrations are analogous to those for the case of one-band superconductors.
For strongly anisotropic systems (the bands are different), the concentration
dependences of $\Delta_{nm}$ and $\mu$ can differ from the case of one-band
superconductors. This is because, in addition to the explicit dependence on the
carrier concentration, a nonexplicit dependence exists through the ratios
$\Delta_{11} /\Delta_{22},\,\Delta_{12} /\Delta_{22}$, etc.

3. The difference in the thermodynamic potentials (62) shows that the
considered condensed state is  advantageous at $\Delta_{nm} \not = 0$. The
advantage of this state increases as the number of order parametrs and,
consequetly, the namber of intraelectron constants, increases.

To make the results more illustrative, two simpler cases - the limiting case of
the Moskalenko model [2], where $\Delta_{11}, \Delta_{22} \not = 0,
\Delta_{12} = 0$, as well as the case where $\Delta_{11} = \Delta_{22} = 0,
\Delta_{12} \not =0$ - were considered. We have show that  overlapping of the
energy bands facilitate superconductivity and increase the critical
concentration of the carriers at which a transition from the BCS scenario to
the local-pair condensate scenario occurs. In the case of different bands
(especially, for strong anisotropy), the dependence of $\Delta_{nm}$ and $\mu$
on the carrier concentration may be different from the one-band superconductor
case because of their additional implicit dependence on $\tilde n$ through the
ratio $\Delta_1 / \Delta_2$. We also obtained an equation for the bound-state
energy $\varepsilon_b$ on a two-band basis and established the relation
$\varepsilon_b = 2 \mu$.

In section 3.2 is developed the  path integral method for two-band
superconductors and, on this basis, inprove the mean field approximation to
make it applicable for $T \not = 0$.

To do this, we introduced action, wich takes into account the existence of
intra - and interband interactions and additional boson fields. In the equation
for the statidtical sum, we performed the Hubbard-Stratanovich transformation
generalized to the two-band case. Having perfomed the integration over Fermi
fields, and expansion in terms of $\hat \Delta_n$ with an accuracy to quadratic
terms, we came to effective action (66),(67). We connected the calculation of
$\Pi_n (\vec q, \omega)$ with the existence in the system of a bound state with
the bound-state energy $|\varepsilon_b| \gg T$ and performed the expansion of
$\Pi_n (\vec q, \omega)$ in powers  of $\omega,\,q^2$, and
$\left(\varepsilon_b - 2 \mu \right)$. Along the way, we came to statistical
sum for the ideal Bose gas with a renormalized chemical potential $\mu^* =
2 \mu - \varepsilon_b$ and effective mass $M^*$ (75).

We also determined the condensation temperature $T_k$, Eq. (76). In the case
of different energy bands, considerable renormalization connected with the
existence of two energy bands occurs. At the same time, the dependence of $T_k$
on the carrier concentration is explicity determined by the dimension of the
system and is implicitly determined by the dependence of $M^*$ on the carrier
concentration.

The contribution of the residual Bose interaction to the effective action was
also calculated. In analogy to the case of a one-band superconductor, this
contribution is unessential only in three-dimensional systems
\cite{Gorbatsevich_1}. For such systems, the mean field approximation can be
applied, while for systems of reduced dimension, the mean field approximation
appears to be insufficient.

The main results of the section 3.2 are following:

1. On the basis of the functional integration method, the procedure of the
transition from the BCS scenario to the Schafroth one is developed for the
two-band superconductor, when changing the charge carrier density. In the
mean-field approximation the ideal Bose gas with renormalized chemical
potential $\mu^* = 2\mu - \varepsilon_b$ and effective mass $M^*$ is
obtained.These quantities are dettermined on the basis of the two -
band model. The anisotropy of the energy bands plays the critical role in the
renormalization of the effective mass. In particular, at $m_1 > m_2$ the easier
Bose gas $( M^* < m_{1}^{*})$ appears in comparison with the case of a single
band with effective mass $m_1$.

2.In the case of strongly anisotropic two-band systems $(\Delta_1 \not =
\Delta_2)$ the possibility of increasing the temperature of the Bose
condensation of localized pairs $T_k$ in comparison with the case of the
single energy band is shown.

3. The contribution of residual interaction between bosons to the  effective
action depends essentially on the system dimension and has a low value only
for a three-dimensional systems.Therefore for systems with reduced dimension
it is necessary to depart from the mean-field approximation.\\
\begin{center}
\bf{4. Summary}.
\end{center}

High-$T_{C}$ superconductors as was mentioned in Introduction are very
complicated compounds, and it is impossible to take into account all their
features simultaneously. So, the theoretical studies are
performing on the basis of the models, which take into account some
characteristic features of these systems.

The goal of this articl is to make the review of the
papers, in which two peculiarities are taken into consideration
- the overlapping of the energy bands on the Fermi surface and the reduced
(or low) density of the charge carriers.

In the two-band model proposed by prof. Moskalenko \cite{Moskalenko_2}  the
Cooper pairs appears in every energy band and pass as a whole from one
enrgy band to another.

In the systems with the low density of the charge carriers  there is the
necessity to take into consideration  all possible additional pairings
of  electrons from different energy bands \cite{Kochorbe_1} and go beyond the
approximation, in which are taken into account only the diagonal over the
band indices Green functions \cite{Moskalenko_9}. The basic  equations of the
theory of superconductivity in such two-band systems is reduced to the
representation of four energy pseudo-bands. In this case there appears three
energy gaps $\Delta_{11}$, $\Delta_{22}$ and $\Delta_{12}=\Delta _{21}$ and it
is also possible to obtain the values of the temperature of the superconducting
transition $T_c$, characteristic to the  high-$T_C$ materials, even at
the low density of the charge carriers as in the case of the electron
attraction ($\lambda_{nm}>0$), as in the case of their repulsion
($\lambda_{nm}<0$).  The main attention in this paper is devoted to the
dependence of the thermodynamic quantities on the density of charge carriers.
In particular, the behavior of the quantity $T_c$ and the jump of electron
heat capacity ($C_S-C_N$) at the point $T=T_c$ as a function of the density of
the charge carriers is determined by the degree of the overlapping of the
energy bands, of the filling of these bands, relation between the constants
of the electron-electron interaction $\lambda_{nm}$ ($n; m=1-4$) and the
ratio of electron density states $N_1 / N_2$.

As, for example, in the case of strong hybridization $T_c$ and
$(C_S-C_N)$ as  function of the concentration of the charge carriers
are represented by the bell-shaped dependence, which is
observed in the experiment in the number  of the oxide ceramics. The
represented theory allows to obtain as small $(C_s-C_N)/C_N<1.43$,  as
large $(C_s-C_N)/C_N>1.43$ values for the relative jamp of the electron heat
capacity.  This picture is observed in the high-temperature materials
\cite{Junod}-\cite{Akis_2}. These studies have fulfiled in the
representation of the BCS superconducting Cooper pairs. At decreasing  the
density of charge carriers at the point $\mu =0$ there takes place the
crossover BCS state-Bose condensation of the localized pairs.  In the state of
the deep Bose-condensation, when $\mu < 0$ and the relation
$\Delta_{n}^{2}/\mu^2 \ll 1$ takes place at $T=0$, we obtain that analytical
expressions for the quantities $\Delta_n$ and $\mu$ contain additional
inexplicit dependence on the density of charge carriers in comparison with the
single-band system, due to the consideration of the overlapping of the energy
bands on the Fermi surface.

Given in the article the method of the functional integration applied to the
two-band system with low density of the charge carriers demonstrates the
transition procedure from the Fermi to the Bose elementary excitations at $ T
\not = 0$.  We have obtain ideal Bose-gas with the effective mass $M^{*}$ and
chemical potential $\mu^*=2\mu- \varepsilon_b$ ($\varepsilon_b$ is the energy
of the bound two-particle state). These quantities are determined on the
two-band basis.

The important results of the given work are: the overlapping of the energy
bands on the Fermi surface promote to the
appearing of the superconductivity, to explain some
experimental data in the investigations of the thermodynamical properties of
the system,  intensity the number of the effects,
characteristical to the systems with lowered density of the charge carriers,
and assist their experimental confirmation. These effects, for example, are
the appearing  of the kink in the temperature dependence of the chemical
potential $\mu(T)$ in the point $T=T_c$, and also  crossover the state
BCS - Bose condensation of the local pairs. The theory proposed in the paper
can be applied to the oxide ceramics and also to the high-temperature
composition $MgB_2$ and other compositions, in which takes place the
overlapping of the energy bands on the Fermi surface and there is possibility
to change the concentration of the charge carriers.

In addition to the overlapping of the energy  band, an important factor of the
band structure of high-temperature materials is the existence of high symmetry
points in the momentum space that lead to singularies in the electron densitty
of states (Van Hove singularities), as well as to topological electron
transitions. The behavior of these singularities dependends, to a large extent,
on the dimension of the system.

Numerouse theoretical and experimental investigations have confirmed that in
quasi two-dimensional systems (the majority of high-temperature materials are
just such system), a logarithmic divergence of the density of states occurs in
the vicinity of half filling of the energy band at the point where the Fermi
surface intersects the Brillouin zone boundary. This corresponds to the change
from electron to hole topology on the Fermi surface.

This singularity gives rise to high superconducting transition temperatures
$T_c$ and anomalies in the isotope effect and specific heat(see the review in
\cite{Markiewicz}), as well as providing an explanation for experimental
data on the temperature dependence of the resistance and thermal EMF in high-
$T_c$ materials \cite{Palistrant_11}, etc. Photoemission experiments have shown
that logaritmic divergence is a characteristic of the $La_{2 -x} Sr_x CuO_4$
compound  \cite{Gofron}. At the same time, the densities of
electron states may have stronger divergences, such as $E^{-1 / 4}$ and $E^{-1
/ 2}$, in $Y Ba_2 Cu_3 O_{7 - \delta}$ and $BiSrCaCu_2O_8$ compounds,
respectively ($E$ is energy counted from the Van Hove singularity), and
correspond to the so-called extented Van Hove sigularities
\cite{Abricosov_2}.

In a number of articlies, several attempts have been made to explain high-$T_c$
by accounting only for these singularities of the density of electron states
and assuming that the pair interaction is universal for all compounds. In the
weak-coupling approximation, accounting for these singularities results in a
larger increase in the superconducting transition temperature $T_c$ in
comparison with the case of logarithmic divergence . In the
case of strong coupling, the picture changes, leading to an essential decrease
in the role of the extented Van Hove singularities that are spread apart due to
the inelastic scattering of electrons \cite{Radtke}.

It should be noted that we investigated the influence of the density of the
state singularities of type $E^{- 1 / 2}$ on the thermodynamic and kinetic
proterties of superconductors long before the discovery  of high-$T_c$
superconductivity \cite{Palistrant_12}, \cite{Palistrant_13}. In
particular, we showed that the scattering of electrons on the impurity
potential decreased the role of that singularity in the determination of $T_c$.

In studying the properties of high-temperature materials, it is probably
insufficient to consider only the existence of extended Van Hove singularities.
It is necessary to use a more complex approach that accounts for energy band
overlapping,reduced corrier concentrations, nonadiabacity and the other factors
characteristic of high-temperature materials, especially becous the two-band
model is capable of explaining a large amount of experimental a great number
\cite{Moskalenko_5} - \cite{Moskalenko_8}, \cite{Kochorbe_1},
\cite{Palistrant_14}, \cite{Palistrant_15}, in pure and doped high - $T_c$
superconductors including $MgB_2$ \cite{Todor}.

The discovery of high $T_c$ superconductivity in $MgB_2$ and in some other new
materials with the lowered density of the charge curriers persuade us that
given reviw article is well useful for theoretical and experimental
investigations.

We note that there are many authors at present which carry out the theoretical
investigations of the physical properties of $MgB_2$ on the bases of two - band
model (see, for example, in book \cite{Rare}, also in
\cite{Bud'ko}-\cite{Choi_3},\cite{Rodriguez}, \cite{Lagos} and al. and
references in them).

The information about theory investigations of thermodynamic and kinetic
properties of many-band superconductors, elaborated long before discovery of
superconductivity in $MgB_2$, are in arXiv: Cond. Mat.
\cite{Palistrant_1},\cite{Moskalenko_8}, \cite{Kon} and in this review article.
The comparison of this theory with the investigations of last years permits to
physisists, working with above-mentioned problems, to do conclusions about
situation in given field.

{\bf Acknowledgments.}

The author gratefully acknowledges Prof. to V.A. Moskalenko, for his interest
in this problem and is thankful to Dr. F.G.Kochorbe for long-term  collaboration
and to Prof. T.Mishonov for his active position in the  recognition of the
priority accomplishment of Moldavian physicists to the development of the
theory of multi-band superconductors.


\begin{thebibliography}{98}

\bibitem{Moskalenko_1} V. A. Moskalenko, P. Entel, and D. F. Digor,
{\it Phys. Rev.B}, {\bf 59}, 619 (1999).

\bibitem{Moskalenko_2} V. A. Moskalenko,  {\it Fiz. Met.Metalloved};
{\bf 8}, 503 (1959); {\it Phys. Met. and Metallog.} {\bf 8}, 25 (1959) 

\bibitem{Suhl} H. Suhl, B. T. Matthias, and L. R. Walker,
{\it Phys. Rev. Lett.} {\bf 3}, 552 (1959). 

\bibitem{Krakauer} H. Krakauer and E. Pickett,  {\it ibid}, {\bf 60}, 1665
(1988). 

\bibitem{Herman} J. F. Herman, R. V. Kasowski, and W. G. Hsu,  {\it
Phys. Rev. B.}, {\bf 36}, 6904 (1987).

\bibitem{Baranov} M. A. Baranov and Yu. M. Kagan, {\it Zh. Eksp. Teor. Fiz.},
{\bf 102}, No. 1, 313 (1992) [{\it Sov. Phys. JETP} {\bf 75}, 165 (1992)]. 

\bibitem{Palistrant_1} M. E. Palistrant, arXiv :{\it cond - mat / 0305496},
26 june 2003.

\bibitem{Lee_1} D. H. Lee and J. Ihm, {\it Sol. State Commun.}, {\bf 62}, 81
(1987).

\bibitem{Moskalenko_3} V. A. Moskalenko, M. E. Palistrant, and V. M. Vakalyuk,
{\it Mechanisms of High - Temperature Superconductivity} [in Russian],
{\it Joint Inst. for Nucl. Research}, Dubna, (1988), p. 34.

\bibitem{Moskalenko_4} V. A. Moskalenko, M. E. Palistrant, and V. M. Vakalyuk,
{\it Fiz. Nizk. Temp.}, {\bf 15}, 378 (1989) [{\it Sov. J. Low - Temp. Phys}].
{\bf 15} (1989).

\bibitem{Galaiko} V. P. Galaiko, E. V. Bezuglyi, and V. S. Shumeiko {\it
ibid.} {\bf 13},1301 (1987) 

\bibitem{Konsin} P. Konsin, N. Kristoffel, and T. Ord, {\it Phys. Lett. Ser.A}
{\bf 129}, 399 (1988)

\bibitem{Volovik} E. G. Volovik, {\it Pism'ma v Zh. Eksp. Teor. Fiz.}, {\bf
49}, 185 (1989) [{\it JETP Lett.} {\bf 49}, 214 (1989)].

\bibitem{Moskalenko_5} V. A. Moskalenko, M. E. Palistrant, V. M. Vakalyuk,
and I. V. Padure, {\it Solid. St. Commun.}, {\bf 69}, 747 (1989).

\bibitem{Moskalenko_6} V. A. Moskalenko, M. E. Palistrant, and V. M. Vakalyuk,
{\it 10th Internat. Symp. on the Jahn-Teller Effect},Kishinev (1989), p.88.

\bibitem{Moskalenko_7} V. A. Moskalenko, M. E. Palistrant, and V. M. Vakalyuk,
{\it Procced. Eigh International Conference on Ternary and Multinary
Compounds, Kishinev.}, p. 57, (1990).


\bibitem{Palistrant_2} M. E. Palistrant and F. G. Kochorbe {\it Physica C},
{\bf 194}, 351 (1992).

\bibitem{Palistrant_3} M. E. Palistrant and Vakalyik, {\it Sverhprovodimost':
Fiz. Khim. Tekh.}, {\bf 3}, 557 (1990).

\bibitem{Palistrant_4} M. E. Palistrant and F. G. Kochorbe {\it Quantum-Field
Methods of Investigating High} - {\it Temperature Supercanductors and Disordered
Systems [in Russian] Shtiintsa}, Kishinev (1992).


\bibitem{Kalalb_1} M. G. Kalalb, F. G. Kochorbe, and M. E. Palistrant {\it
Teor. Mat. Fiz.}, {\bf 91}, 483 (1992).

\bibitem{Palistrant_5} M. E. Palistrant and M. G. Kalalb, {\it Izvest. Akad.
Nauk, Resp. Moldova}, {\bf 1(7)},70 (1992).

\bibitem{Moskalenko_8} V. A. Moskalenko, M. E. Palistrant, and V. M. Vakalyuk,
{\it Usp. Fiz. Nauk}, {\bf 161}, 155 (1991) {\it [Sov. Phys. Usp.}{\bf 34},
717 (1991), arXiv:{\it cond-mat / 03099671}

\bibitem{Moskalenko_9} V. A. Moskalenko, L. Z. Kon, and M. E. Palistrant,
{\it Low-Temperature Properties of Metals with Band-Spectrum Singularities}
[in Russian], Shtiintsa, Kishinev (1989).

\bibitem{Moskalenko_10} V. A. Moskalenko and M. E. Palistrant,
{\it Statistical Physics and Quantum Field Theory} [in Russian],Nauka, Moscow
(1973), p. 262.

\bibitem{Moskalenko_11} V. A. Moskalenko, {\it Electromagnetic and Kinetic
Properties of Superconducting Alloys with Overlapping Energy Bands} [in
Russian],Shtiintsa, Kishinev (1976).

\bibitem{Kresin_1} V. Kresin and S.A. Wolf, {\it Phys. Rev B}, {\bf 41}, 4278
(1990); {\it Physica C} {\bf 169}, 476 (1990)

\bibitem{Hirch} J. E. Hirch and F. Marsiglio {\it Phys. Rev B}, {\bf 43}, 424
(1991).

\bibitem{Nagamatsu} J. Nagamatsu, N. Nakagawa, T. Muranaka, Y. Zenitani and J.
Akimutsu, {\it Nature (London)}, {\bf 410}, 63 (2001).

\bibitem{Bud'ko} S. L. Bud'ko, G. Lapertot, C. Petrovich at al.  , {\it
Phys. Rev. Lett.}, {\bf 86},1877 (2001).

\bibitem{Jiu_1} A. Y. Liu, I. I. Mazin, and I. Kartus,{\it Phys. Rev. Lett.}
{\bf 87}, 087005 (2001).

\bibitem{Kong} Y. Kong, O. V. Dolgov, O. Jepsen and O. K. Andersen, {\it
Phys. Rev. B}, {\bf 64}, 020501-1  (2001).

\bibitem{Choi_3} H. J. Choi, D. Roundy, H. Sun et al., {\it Nature},{\bf 418},
758 (2002).

\bibitem{Buzea} C. Buzea and T. Yamashita,{\it Supercond. Sci. Technol.} {\bf
14} R 115 (2001).

\bibitem{Ponomarev} Y. U. Ponomarev, S. A. Kuzmichev at al, (unpublished).

\bibitem{Mishonov_1} T. Mishonov and E. Penev, arxiv: {\it cond-mat / 0206118 V
2} 24 june (2002); {\it International Journal of Modern Physics B}.

\bibitem{Mishonov_2} T. Mishonov, S. L. Drechsler and E. Penev,{\it arxiv:
cond-mat / 0209192 V 1} 8 sep. (2002).

\bibitem{Mishonov_3} T. Mishonov, E. Penev, J. O. Indekeu and V. I.
Pokrovsky arxiv:{\it  cond-mat / 0209342 V 2} 17 Mar. (2003);{\it Phys. Rev. B}
{\bf 68}, 104517 (2003).


\bibitem{Moskalenko_12} V. A. Moskalenko, and M. E. Palistrant,
{\it Dokl. Akad. Nauk SSSR} {\bf 162}, 539 (1965) [Sov. Phys. Dokl].

\bibitem{Moskalenko_13} V. A. Moskalenko and M. E. Palistrant,
{\it Zh. Eksp. Teor. Fiz.} {\bf 49}, 770 (1965) {\it [Sov. Phys. JETP}{\bf
22}, 536 (1965)].

\bibitem{Van} V. Van der Marel, {\it Physica C},  {\bf 165}, 35 (1990).

\bibitem{Schaffroth} M. R. Schafroth , {\it Phys.Rev.},  {\bf 111}, 72
(1958).

\bibitem{Migake} K. Migake, {\it Prog. Teor. Phys.},{\bf 69}, 1784 (1983).

\bibitem{Randeria} M. Randeria, J. M. Duan, and L. I Shich, {\it Phys. Rev.
Lett.}, {\bf 62}, 981 (1989); {\it Phys. Rev. B.}, {\bf 41}, 372 (1990).

\bibitem{Michas} R. Michas, J. Rannienger, and S. Robaszkevicz, {\it Rev.
Mod.  Phys.}, {\bf 62}, 113 (1990).

\bibitem{Leggett_2} A. J. Leggett,{\it Modern Trends in the Theory of Condensed
Matter}, {\bf 115}, 13 (1980).

\bibitem{Gorbatsevich_1} A. A. Gorbatsevich and I. V. Tokatly, {\it JETP},
{\bf 76}, 347 (1993).

\bibitem{Gordar} E. V. Gorbar, V. P. Gusynin, and V. M. Loktev, {\it
Sverhprovodimost': Fiz. Khim. Tekh.,}, {\bf 6}, 483 (1993); {\it Fiz. Nizk.
Temp.,} {\bf 19}, 1171 (1993).

\bibitem{Loktev}  V.M. Loktev, R. M. Quick, S. G. Sharapov,  {\it Physics
Reports},{\bf 349}, 1 - 123 (2001).

\bibitem{Kochorbe_1} F. G. Kochorbe, M. E. Palistrant, {\it Zh. Eksp. Teor.
Fiz.}, {\bf 104}, 3084 (1993);{\it JETP}, {\bf 77} 442 (1993); {\it [ Teor.
Mat.  Fiz.} {\bf 96}, 459 (1993); {\it Theoret. Mathemat. Phys.}{\bf 96},1083
(1993);
{\it [ 4th Internat.  Conference M2S - HTS IV Grenoble (France), FR-PS}, 494
(1994).

\bibitem{Abricosov_1} A. A. Abrikosov, L. P. Gor'kov and I. E. Dzyaloshinskii,
{\it Quantum Field Theoretical Methods in Statistical Physics};{\it [Prentice
- Hall, Englewood Cliffs}, N 7; (1963); {\it Moscow, Nauka} (1962)].

\bibitem{Robashkiewicz} S. Robashkiewicz, R. Mikhas, and G. A. Chao, {\it
Phys.Rev.  B}, {\bf 26}, 3915 (1982).

\bibitem{Cardona} M. Cardona and L. Ley, eds., {\it Photoemission in Solids,
Springer} (1978).

\bibitem{Hongshum},H. Hongshum,  X. Zhan, et al, {\it Physica C},
{\bf 172},71 (1990)

\bibitem{Junod} A. Junod, D. Eckert, at al.,{\it ibid} {\bf 159}, 215
(1989).

\bibitem{Loram} J. W. Loram and K. A. Mirza,  {\it ibid},{\bf 153 - 155},
1020 (1988).

\bibitem{Akis_1} R. Akis, F. Morsiglio, and J. P. Carbotte, {\it Phys. Rev. B},
{\bf 39},2722 (1989)

\bibitem{Akis_2} R. Akis and J. P. Carbotte, {\it Physica C},
{\bf 159}, 395 (1989)

\bibitem{Mishonov}T. Mishonov, S. Drechsler and E. Penev, {\it Modern Physics
letter B}, {\bf 17}, 755 (2003)..

\bibitem{Rodriguez} J. J. Rodriguez-Nunez and A. A. Schmidt, {\it Phys. Rev.B
B}, {\bf 68}, 224512 (2003).

\bibitem{Lagos} R. E. Lagos and G. G. Cabrera, {\it Brazilian Juornal of
Physics}, {\bf 33}, 713 (2003).

\bibitem{Jemma}S. Jemma Balaselvi; A. B. Bharthi at al, {\it arXiv: cond-mat
/ 0209200 \,; 0303022}.

\bibitem{Kazakov} S. M. Kazakov, J. Karpinski at al, {\it arXiv: cond-mat
/ 0304686 \,; 0303022}.

\bibitem{Palistrant_5a} M. E. Palistrant, {\it Low Temperature Physics} {\bf
26}, 799 (2000).

\bibitem{Palistrant_6} M. E. Palistrant, {\it Teoret. Matemat. Fiz.},
{\bf 95},101 (1993).

\bibitem{Palistrant_7} M. E. Palistrant, {\it Teoret. Matemat. Fiz.},
{\bf 105}, 491 (1995) ; {\it Teoret. and Mathem. Phys.} {\bf 105}, 1593
(1995).

\bibitem{Palistrant_8} M. E. Palistrant, V. M. Vackalyuk, M. G. Calalb,  {\it
Physika C}, {\bf 208}, 170 (1993).

\bibitem{Palistrant_9} M. E. Palistrant, {\it Teoret. Matemat. Fiz.},
{\bf 109}, 137 (1996) ; {\it Teoret. and Mathem. Phys.} {\bf 109}, 1352
(1996).

\bibitem{Palistrant_10} M. E. Palistrant, {\it J. of Superconductivity},
{\bf 10}, 19 (1997).

\bibitem{Markiewicz} R. S. Markiewicz, {\it Int. J. Mod. Phys.},
{\bf 5}, 2037 (1991).

\bibitem{Palistrant_11} M. E. Palistrant and F. G. Kochorbe, {\it Izv. Akad.
Nauk Resp. Moldova}, 2(5), {\bf 7} (1991).

\bibitem{Gofron} K. Gofron , J. C. Campuizano, and H. Ding at al.,{\it Phys.
Chem. Solids} {\bf 54} 1193 (1993).

\bibitem{Abricosov_2} A. A. Abrikosov, J. C. Compuizano and  H. Gofron,
{\it Physica C}, {\bf 214}, 73 (1993).

\bibitem{Radtke} R. J. Radtke and M. R. Norman, {\it Phys. Rev. B} {\bf 50},
9594 (1994).

\bibitem{Palistrant_12} M. E. Palistrant and Trifan, {\it Fiz. Nizk. Temp.,}
{\bf 3}, 241 (1977); {\bf 3}, 976 (1977).

\bibitem{Palistrant_13} M. E. Palistrant and O. P. Bezzub, {\it Fiz. Nizk.
Temp.,} {\bf 6}, 1146 (1980); {\bf 9}, 357 (1983).

\bibitem{Palistrant_14} M. E. Palistrant, {\it Teor. Matemat. Fiz.}{\bf 111},
289 (1997), {\it Theoretical Mathem. Phys.} {\bf 111}, 621 (1997).

\bibitem{Palistrant_15} M. E. Palistrant and F. G. Kochorbe, {\it Fiz. Nizk.
Temp.}{\bf 26}, 1077 (2000); {\it Low Temp. Phys.} {\bf 26}, 799 (2000).

\bibitem{Todor} T. M. Mishonov, V. L. Pokrovsky and H. Wei, arXiv: {\it cond.
mat. / 0312210}, Dec. 2003.

\bibitem{Rare} Rare Earth Transition Metal Borocarbides (Nitrides):
Superconducting, Magnetic and Normal State Properties (edited by Karl-Hartmut
Muller and Vladimir Narozhnyi) Nato Sciences Series II. Mathematics, Physics
and Chemistry - Vol. 14.

\bibitem{Kon} L. Z. Kon, arXiv: {\it cond. mat. / 0309707}
\end{thebibliography}
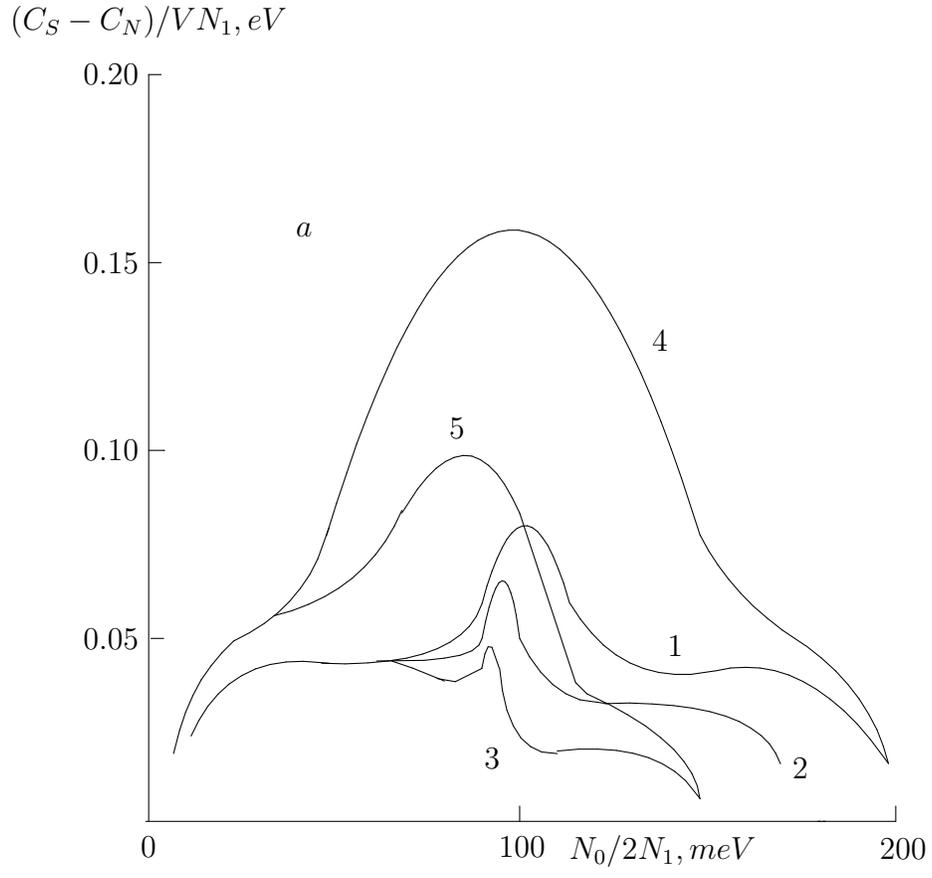
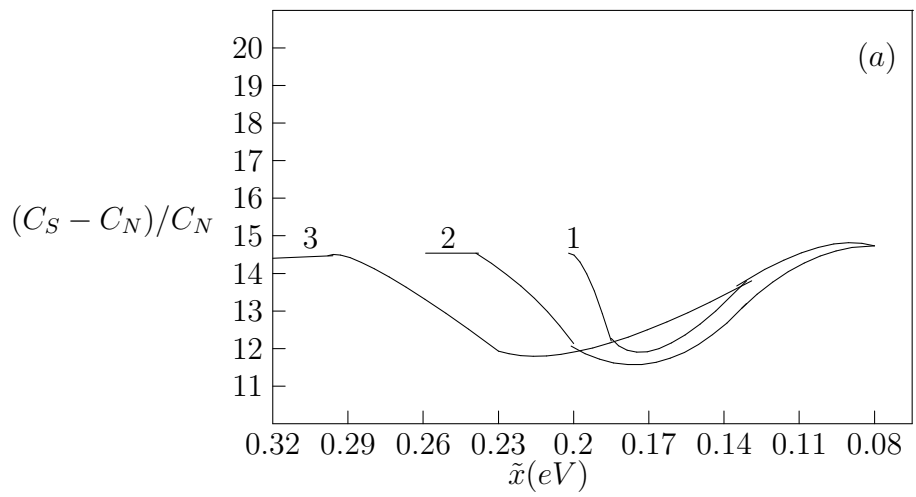
\end{document}